\documentclass{article}

\usepackage{arxiv}

\newcommand{\diag}{\operatornamewithlimits{diag}}

\newtheoremstyle{preprintstyle}
  {10pt plus 2pt minus 2pt}
  {10pt plus 2pt minus 2pt}
  {\itshape}
  {}
  {\color{secondary}\bfseries}
  { - }
  {4pt}
  {}

\theoremstyle{preprintstyle}

\newtheorem{problem}{\textbf{Problem}}

\newtheorem{assumption}{\textbf{Assumption}}

\title{Periodic Motion Optimization for an Underactuated Mechanical System through Synergistic Structure-Control Design}

\author{Andrea~Tilli
		\thanks{Andrea Tilli, Alessandro Bosso, Elena Ruggiano, and Alessandro Samor\`i are within the research group Advanced Control and Technologies for Enhanced Mechatronics and Automation (\href{https://dei.unibo.it/en/research/research-groups/actema}{\color{secondary} ACTEMA}) at the Department of Electrical, Electronic and Information Engineering (\href{https://dei.unibo.it/en}{\color{secondary} DEI}), University of Bologna, Viale Risorgimento 2, 40136 Bologna, Italy.
		\newline Email: [\href{mailto:andrea.tilli@unibo.it}{andrea.tilli},
		\href{mailto:alessandro.bosso3@unibo.it}{alessandro.bosso3},
		\href{mailto:elena.ruggiano2@unibo.it}{elena.ruggiano2},
		\href{mailto:alessandro.samori4@unibo.it}{alessandro.samori4}]@unibo.it
		 }
		\And Alessandro~Bosso \footnotemark[1]
		\And Elena~Ruggiano \footnotemark[1]
		\And Alessandro~Samor\`i \footnotemark[1]
}
\copyrightTransfer{\vskip -0.3in}

\headerAuthor{A. Tilli, A. Bosso, E. Ruggiano, and A. Samor\`i}

\begin{document}
\maketitle

\begin{abstract}
In this work, we present the integrated structure-control design of a 2-DOF underactuated mechanical system, aiming to achieve a periodic motion of the end-effector.
The desired behavior is generated via input-output linearization, followed by structural optimization of the zero dynamics.
Inspired by recent works on the control-oriented design of multibody systems, we define an optimization problem based on the simulation of the system's response.
In particular, relevant model parameters are used to match the reference with a specific orbit of the zero dynamics, while also penalizing the input energy.
For the considered application, the selected parameters are related to the mechanism's elasticities and mass distribution.
Notably, we show that it is possible to reach a desirable trade-off between mass reduction and periodic motion accuracy.
With an optimal zero dynamics response available, the control scheme can be completed with established orbital stabilization techniques, ensuring a robust oscillating behavior.
\end{abstract}


\section{Introduction}

Optimization of complex mechanical systems is nowadays a topic of ever-growing importance.
Indeed, critical challenges arise in several civil and industrial fields, involving both dynamic performance and reliability.
Although the contraposition of these objectives is ubiquitous in engineering problems, the typical approaches often result in suboptimal solutions.
In particular, a common strategy to ensure robustness in industrial applications, especially in high-end manufacturing systems, is to oversize the components.
However, oversizing is designed for conditions far beyond the nominal behavior, thus inevitably leading to higher costs and, overall, reduced sustainability.
In this respect, it is natural to wonder if oversizing can be mitigated (or avoided) by synergistically exploiting the inherent internal features of the designed structure.
\par Regarding this challenging problem, it is fundamental to leverage the vast literature on multibody systems, where the fields of modeling, design optimization, and control are widely documented \cite{wasfy2003computational}.
In this context, structural optimization based on the simulation of the multibody dynamics has received particular attention \cite{tromme2018system}.
Notably, the topological optimization of multidomain systems follows similar principles \cite{ma2006multidomain}; therefore, it can be exploited not only to enhance the mechanism behavior but also to integrate sensors and actuators within.
Indeed, optimization should not be restricted to purely mechanical components, but it should aim at an increased blending of all functional elements.
This idea represents one of the central stepping stones towards an authentic mechatronic design.
In this respect, we refer to \cite{trease2009design} for a recent approach for the organic incorporation of sensing and actuation technologies within the mechanical structure.
\par Specializing the discussion to the development of reliable industrial servomechanisms, the focus of this work is on the maximization of the system's efficiency, aiming at lightweight structures that require low control effort.
In particular, the main objective is to match the desired behavior with the natural response of the system, i.e., to shape the system's resonant behavior.
From the control viewpoint, such an operating condition is intuitively optimal in terms of efficiency but potentially fragile.
It becomes conspicuous that the structural design can no longer be decoupled from the control architecture.
In fact, appropriate input signals are needed to impose the realized resonances.
Then, robust feedback laws must be introduced to avoid disruptive phenomena that might occur, e.g., if the system deviates significantly from the nominal trajectories.
\par The natural framework where such challenges can be appropriately formulated is that of Euler-Lagrange (EL) systems.
Of particular interest, in this context, we find underactuated EL systems \cite{de2002underactuated}, \cite{liu2013survey}, \cite{olfati2001nonlinear}.
Several studies have been devoted to problems of stabilization and tracking of underactuated EL systems, including input-output linearization \cite{spong1994partial} and passivity-based techniques \cite{ortega2002stabilization}.
Additionally, a goal of significant practical importance is the achievement of periodic trajectories.
Among the strategies that address this problem, a notable example is given by Virtual Holonomic Constraints (VHC) \cite{shiriaev2005constructive}, \cite{mohammadi2018dynamic}, which involve the orbital stabilization of the reference.
\par The above control-oriented works, however, draw formal results considering only a fixed structure: as a consequence, full optimization cannot be ensured a priori.
In this respect, a crucial contribution towards a fully synergistic approach is found in \cite{seifried2014dynamics}, where structural optimization is tightly coupled with the control design.
There, the general intuition regarding the natural behavior of the system is translated into the shaping of the zero dynamics.
From this description, it is then possible to generate feedforward actions through the right-inversion of the dynamics, which can be further simplified by ensuring, through design, that the system is minimum phase.
Similarly, in \cite{bastos2019synergistic} the authors optimize the internal dynamics to address the problem of trajectory tracking, with the aim to minimize both the control action and the system vibrations.

\subsection{Our Contribution}
As anticipated in the previous discussion, the interest of this work is to further develop the control-oriented structural optimization of underactuated EL systems by formally considering the design of oscillating (resonant) zero dynamics.
This element is of a particular novelty since the works above generally aim at the minimization of vibrations.
On the other hand, considering such a behavior offers intriguing versatility.
Indeed, the optimization of the internal trajectories becomes instrumental in extending the limitations of underactuated systems.
From this perspective, the number of outputs for control can be higher than the actuated Degrees of Freedom (DOF).
\par Here, the codesign problem is addressed through a meaningful case of study, given by a 2-DOF underactuated EL system.
The considered mechanism is based on a four-bar linkage, already subject of previous control design \cite{bosso2019constrained}, augmented with an additional DOF.
Such a complication can be considered as a starting point for the development of a flexible end-effector for industrial applications.
Given a target trajectory in the workspace and the geometric properties of the mechanism, we decouple the problem into zero dynamics optimization and orbital stabilization.
\par Regarding the first part of the procedure, we select some relevant system parameters, in order to match the resulting trajectories with the reference, minimize the control effort of the feedforward actions, and ensure simple stability (i.e., a center) of the zero dynamics.
With the latter property, typical conditions required for orbital stabilization can be imposed by design.
We remark that these operations do not involve output relocation, as proposed, e.g., in \cite{seifried2014dynamics}, \cite{sreenath2011compliant}.
The results are thus shown to be a tunable trade-off between motion accuracy, minimization of the control effort, and reduction of the weight.
Finally, orbital stabilization is solved through the VHC approach, based in particular on the procedure found in \cite{shiriaev2005constructive}.
\par The paper is organized as follows.
In Section \ref{sec:problem_and_strategy}, we present the considered case of study, formally introduce the synergistic design problem, then provide the main steps of the proposed strategy.
Section \ref{sec:optimization} is dedicated to analyzing the zero dynamics optimization.
In particular, two design choices and different algorithms are proposed for this step.
Then, in Section \ref{sec:control} we provide details on the controller used to achieve robust orbital stabilization.
Finally, in Section \ref{sec:conclusions} we present some conclusive remarks and possible future directions.

\subsection{Notation}
In this work, we indicate with $(\cdot)^T$ the transpose of real-valued matrices, while $I_n$ represents the identity matrix of size $n$.
For simplicity, given a pair of column vectors, $u$, $v$, the notation $(u, v)$ is often used to denote the concatenated vector $(u^T\; v^T)^T$.
Furthermore, the time argument of signals is omitted when clear from the context.

\section{A Synergistic Design for Periodic Motion Planning}\label{sec:problem_and_strategy}
\begin{figure}[b!]
	\centering
	\psfragscanon
	
	\vspace{3pt}
	
	\psfrag{a} [B][B][0.7][0]{$q_1$}
	\psfrag{b} [B][B][0.7][0]{$q_2$}
	\psfrag{c} [B][B][0.7][0]{$q_3$}
	\psfrag{d} [B][B][0.7][0]{$q_4$}
	\psfrag{l}  [B][B][0.7][0]{$l_1$}
	\psfrag{m}[B][B][0.7][0]{$l_2$}
	\psfrag{n} [B][B][0.7][0]{$l_3$}
	\psfrag{o} [B][B][0.7][0]{$l_4$}
	\psfrag{u} [B][B][0.7][0]{$u$}
	\psfrag{h} [B][B][0.7][0]{$h(q) = \begin{pmatrix}x\\ y\end{pmatrix}$}
	\psfrag{x} [B][B][0.7][0]{$e_x$}
	\psfrag{y} [B][B][0.7][0]{$e_y$}
	\includegraphics[clip = true, width = 0.45\textwidth]{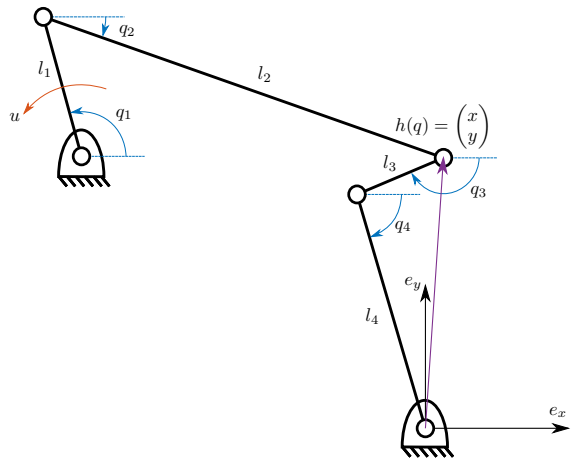}
	
	\caption{Geometric scheme of the 2-DOF mechanism used as a case of study. The joint coordinates are positive when associated with a counterclockwise rotation (hence $q_2$, $q_3$, $q_4$ are negative in the figure), while the pair $e_x$, $e_y$ is the reference frame of the workspace. In such a frame, the position of the end-effector is indicated in violet. For simplicity, we omit to represent explicitly the springs later employed for structural optimization (see Figure \ref{fig:springs_and_link_structure}-Left).}
	\label{fig:mechanism}
\end{figure}

The case of study that we consider is presented in Figure \ref{fig:mechanism}, and is given by a multibody system with kinematic loop constraints.
As anticipated, this structure represents a modification of a four-bar linkage, used as a case of study in \cite{bosso2019constrained}.
The proposed augmentation with an additional DOF is aimed at enhancing the precision of horizontal motion laws of the end-effector, a particularly desirable feature in the context of manufacturing systems.
For simplicity, we consider fixed lengths of the links, while the decision variables for optimization are chosen to only affect the dynamic response.
Future works will focus on optimizing the kinematics as well.
\par The first part of this section is dedicated to the general formulation of the codesign problem, introducing the required mathematical formalism.
Then, we proceed with the main concepts exploited to achieve a horizontal periodic motion.

\subsection{Problem Statement}

Firstly, the dynamics of the considered mechanism is described as the following differential-algebraic equation (DAE):
\begin{equation}\label{eq:el_dae}
\begin{split}
&M(p, q)\ddot{q} + C(p, q, \dot{q})\dot{q} + G(p, q) = bu + \left[\frac{\partial \phi}{\partial q}(q)\right]^T  \lambda\\
&\phi(q) = 0,
\end{split}
\end{equation}
where $q \coloneqq (q_1, q_2, q_3, q_4) \in \mathbb{R}^4$ is the vector of generalized coordinates, $\lambda \in \mathbb{R}^2$ are the Lagrange multipliers accounting for the kinematic loop constraint $\phi(q) = 0$, while $p \in \mathcal{P} \subset \mathbb{R}^d$ is a vector of parameters affecting the dynamic response of the system, belonging for simplicity to a convex compact set $\mathcal{P}$.
We do not specify $p$ as we will later provide two choices, depending on desired the level of complexity/optimality.
In addition, $M: \mathbb{R}^d \times \mathbb{R}^4 \to \mathbb{R}^{4 \times 4}$, $C: \mathbb{R}^d \times \mathbb{R}^4 \times \mathbb{R}^4 \to \mathbb{R}^{4 \times 4}$, $G: \mathbb{R}^d\times\mathbb{R}^4 \to \mathbb{R}^4$, $\phi: \mathbb{R}^4 \to \mathbb{R}^2$ are smooth maps in their respective arguments.
Finally, since only the prime mover coordinate $q_1$ is actuated, it holds $b = (1\; 0\; 0\; 0)^T$.
In addition to the above DAE, consider as output for control the position of an end-effector in the workspace (see Figure \ref{fig:mechanism}):
\begin{equation}\label{eq:xy_output}
\chi = h(q) = \begin{pmatrix}
x\\
y
\end{pmatrix},
\end{equation}
where $h: \mathbb{R}^4 \to \mathbb{R}^2$ is a smooth map.
To this output, we associate a periodic target trajectory $r$, with period $T_r$.
In particular, to achieve a horizontal motion, we choose
\begin{equation}\label{eq:xy_ref}
r(t) = \begin{pmatrix}
r_x(t)\\
\bar{r}_y
\end{pmatrix},
\end{equation}
with $\bar{r}_y$ a constant scalar.
\par Even if the above system is characterized by a single input, the trajectory $r$ is a 2-dimensional signal.
In this context, we will show that a motion corresponding to $r$ is achieved through suitable selection of the parameters $p$, in addition to the design of the control input $u$.
However, since we aim at shaping the internal dynamics of system \eqref{eq:el_dae} through the constant vector $p$, we do not expect to exactly preserve the period of the original reference trajectory.
Therefore, for a meaningful initial investigation, we instead propose to approximate the shape of $r$ (and its derivative) with the behavior of the system, while preserving as much as possible the original period $T_r$.
These concepts are formulated in the following problem, involving both the shaping of the system response and an appropriate stabilization strategy.

\begin{problem}\label{general_problem}
Consider system \eqref{eq:el_dae}, with output for control \eqref{eq:xy_output}.
Let $r$ be a $T_r$-periodic trajectory as in \eqref{eq:xy_ref}, and let $\Gamma_r \coloneqq \{(\xi_1, \xi_2, \xi_3, \xi_4) \in \mathbb{R}^4 : \xi_1 = r_x(t), \xi_2 = \dot{r}_x(t), \xi_3 = \bar{r}_y, \xi_4 = 0, 0 \leq t \leq T_r \}$.
Design:
\begin{itemize}
\item a parameter vector $p \in \mathcal{P}$;
\item a (possibly dynamic) controller of the form
\begin{equation}\label{eq:generic_ctrl}
\begin{split}
u &= \mu(p, q, \dot{q}, \eta, t)\\
\dot{\eta} &= \nu(p, q, \dot{q}, \eta, t),
\end{split}
\end{equation}
with $\eta \in \mathbb{R}^{n_{\text{c}}}$, $n_{\text{c}} \in \mathbb{Z}_{\geq 0}$, and $\mu$, $\nu$ appropriate maps;
\end{itemize}
satisfying the following properties:
\begin{itemize}
\item the closed-loop system \eqref{eq:el_dae}-\eqref{eq:generic_ctrl} is such that there exists a bounded periodic trajectory of the form $(\bar{q}(\cdot), \dot{\bar{q}}(\cdot), \bar{\eta}(\cdot))$, with period $T_q$.
Let $h_{\bar{q}}(t) = h(\bar{q}(t))$, for all $t$, then consider $\Gamma_\chi \coloneqq \{(\xi_1, \xi_2, \xi_3, \xi_4) \in \mathbb{R}^4 : (\xi_1, \xi_3) = h_{\bar{q}}(t), (\xi_2, \xi_4) = \dot{h}_{\bar{q}}(t), 0 \leq t \leq T_q \}$;
\item by minimizing a cost functional related to the parameters, denoted with $J(p)$, the set $\Gamma_\chi$ approximates $\Gamma_r$ as much as possible, while preserving limited energy of the associated steady-state input trajectory $u(\cdot)$;
\item the trajectory $(\bar{q}(\cdot), \dot{\bar{q}}(\cdot), \bar{\eta}(\cdot))$ is uniformly orbitally asymptotically stable for the closed-loop system \eqref{eq:el_dae}-\eqref{eq:generic_ctrl}.
\end{itemize}
\end{problem}
We refer to \cite[Definition 8.2]{khalil_3ed} for some basic notions on orbital stability, involving the application of Lyapunov stability concepts to invariant sets given by periodic orbits.

\subsection{The Proposed Strategy}

\subsubsection{Representation of the Mechanism in Minimal Form}

We begin this discussion by providing a sufficient condition to rewrite system \eqref{eq:el_dae} in minimal form.
This is motivated both for computational simplicity and to highlight the structural features of the strategy.
In this respect, we require the following Assumption.

\begin{assumption}\label{hyp:change_of_coordinates}
Consider system \eqref{eq:el_dae}, with output \eqref{eq:xy_output} and $T_r$-periodic reference $r$ as in \eqref{eq:xy_ref}.
The maps $\phi$ and $h$ are such that:
\begin{itemize}
\item there exists a differentiable map $q_r: \mathbb{R}^2 \to \mathbb{R}^4$ such that $h(q_r(r(t))) = r(t)$, $\phi(q_r(r(t))) = 0$, for all $t \in [0, T_r]$;
\item there exist open sets $\mathcal{U}, \mathcal{V} \subset \mathbb{R}^4$ and a diffeomorphism $\rho: \mathcal{U} \to \mathcal{V}$, satisfying $q = \rho(\chi, \psi)$ (for some $\psi \in \mathbb{R}^2$) and $q_r(r(t)) \in \mathcal{V}$, for all $t \in [0, T_r]$;
\item the matrix
\begin{equation}
\mathcal{J}_\psi \coloneqq \frac{\partial \phi}{\partial q}\frac{\partial \rho}{\partial \psi}(\chi, \psi),
\end{equation}
is non-singular in $(\chi, \psi) = \rho^{-1}(q_r(r(t))$, for all $t \in [0, T_r]$, 
\end{itemize}
\end{assumption}

This way, we can employ standard results to replace the original DAE with a 2-DOF unconstrained system.
In particular, there exists an open set $\mathcal{N} \subset \mathbb{R}^2$, satisfying $r(t) \in \mathcal{N}$, for all $t \in [0, T_r]$, and such that \eqref{eq:el_dae} can be rewritten as
\begin{equation}\label{eq:chi_system}
M_\chi(p, \chi)\ddot{\chi} + C_\chi(p, \chi, \dot{\chi})\dot{\chi} + G_\chi(p, \chi) = B_\chi(\chi)u, \qquad \chi \in \mathcal{N},
\end{equation}
for some smooth maps $M_\chi: \mathbb{R}^d \times \mathbb{R}^2 \to \mathbb{R}^{2 \times 2}$, $C_\chi: \mathbb{R}^d \times \mathbb{R}^2 \times \mathbb{R}^2 \to \mathbb{R}^{2 \times 2}$, $G_\chi: \mathbb{R}^d\times\mathbb{R}^2 \to \mathbb{R}^2$, $B_\chi: \mathbb{R}^2 \to \mathbb{R}^2$.
\par To show this result, it is sufficient to follow, mutatis mutandis, the procedure in \cite[Chapter 2.1.5]{seifried2014dynamics}, which we report for completeness.
By means of the implicit function theorem, applied on the equation $\phi(\rho(\chi, \psi)) = 0$, we can write $\psi = \sigma(\chi)$, defined in a neighborhood $\mathcal{N}$ of $r(t)$.
Note that $\rho(r, \sigma(r)) = q_r(r)$.
Computing the time derivative of $\phi$, it holds:
\begin{equation}
\frac{\partial \phi}{\partial q}\frac{\partial \rho}{\partial \chi}\dot{\chi} + \frac{\partial \phi}{\partial q}\frac{\partial \rho}{\partial \psi}\dot{\psi} =  \mathcal{J}_\chi\dot{\chi} + \mathcal{J}_\psi \dot{\psi} = 0, 
\end{equation}
which yields $\dot{\psi} = -\mathcal{J}_\psi^{-1}\mathcal{J}_\chi\dot{\chi}$.
Similarly, we have $\ddot{\psi} = -\mathcal{J}_\psi^{-1}(\mathcal{J}_\chi\ddot{\chi} + \omega_\psi)$, where $\omega_\psi$ results from the derivation of $\mathcal{J}_\chi$ and $\mathcal{J}_\psi$.
This way we can compute $\dot{q}$, $\ddot{q}$ as:
\begin{equation}
\dot{q} = \underbrace{\begin{pmatrix}\frac{\partial \rho}{\partial \chi}& \frac{\partial \rho}{\partial \psi} \end{pmatrix}}_{\mathcal{J}_\rho} \begin{pmatrix} \dot{\chi} \\ \dot{\psi}\end{pmatrix} = \underbrace{\mathcal{J}_\rho \begin{pmatrix} I_2 \\ -\mathcal{J}_\psi^{-1}\mathcal{J}_\chi\end{pmatrix}}_{\mathcal{J}}\dot{\chi}, \qquad \ddot{q} = \mathcal{J}\ddot{\chi} + \omega_q
\end{equation}
It is immediate to verify that $\mathcal{J}^T$ is a left annihilator of $(\partial \phi/\partial q)^T$, therefore premultiplying \eqref{eq:el_dae} by $\mathcal{J}^T$ cancels the Lagrange multipliers.
Finally, replace $q$ with $\rho(\chi, \sigma(\chi))$ and $\dot{q}$, $\ddot{q}$ with the above expressions to yield the desired model.
\par In general, the proposed minimal form is not globally defined and its solutions may leave the set $\mathcal{N}$.
A crucial step to preserve the validity of model \eqref{eq:chi_system} is then to impose that the geometric path of $h(\bar{q}(\cdot))$ (see Problem \ref{general_problem}) is sufficiently close to that of $r(\cdot)$.
This issue can be approached by penalizing the mismatch between the trajectories in the cost functional $J(q)$.

\subsubsection{Input-Output Linearization and Shaping of the Zero Dynamics}

Exploiting the simpler structure of system \eqref{eq:chi_system}, we can effectively develop our approach.
In particular, we take inspiration from the recent works on control-oriented structural design, which perform optimization only on some form of zero dynamics of the system.
In the same fashion, here we highlight the internal dynamics performing partial feedback linearization.
For this purpose, rewrite system \eqref{eq:chi_system} as:
\begin{equation}
\begin{split}
\ddot{x} &= f_x(p, x, \dot{x}, y, \dot{y}) + g_x(p, x, y)u\\
\ddot{y} &= f_y(p, x, \dot{x}, y, \dot{y}) + g_y(p, x, y)u.
\end{split}
\end{equation}
Let $\bar{r}_y$ be the output of the feedback linearizing controller, and consider the following Assumption.

\begin{assumption}
For all $p \in \mathcal{P}$ and all $(x, y) \in \mathcal{N}$, it holds $g_y(p, x, y) \neq 0$.
\end{assumption}

We can thus assign the input $u$ as
\begin{equation}\label{eq:i_o_lin}
u = -\frac{f_y(p, x, \dot{x}, y, \dot{y})}{g_y(p, x, y)} + \frac{v}{g_y(p, x, y)} = \tau_0(p, x, \dot{x}, y, \dot{y}) + \tau_v(p, x, y)v,
\end{equation}
with $v$ an input that can be used for stabilization.
We obtain the following system:
\begin{equation}\label{eq:partially_lin_system}
\begin{split}
\ddot{x} &= f_x(p, x, \dot{x}, y, \dot{y}) + g_x(p, x, y)\left[\tau_0(p, x, \dot{x}, y, \dot{y}) + \tau_v(p, x, y)v\right]\\
\ddot{y} &= v
\end{split}\qquad e = y - \bar{r}_y,
\end{equation}
where we indicated with $e$ the tracking error, regarded as the output of the system.
Then, the zero dynamics associated with $e = 0$ is a second-order autonomous system, given by
\begin{equation}
\ddot{x} = f_x(p, x, \dot{x}, \bar{r}_y, 0) + g_x(p, x, \bar{r}_y)\tau_0(p, x, \dot{x}, \bar{r}_y, 0).
\end{equation}
Taking advantage of the zero dynamics, our approach for Problem \ref{eq:generic_ctrl} can be divided into two functional steps:
\begin{itemize}
\item an optimization problem is set up to match the reference $r_x$ with the trajectories that the zero dynamics can generate, while penalizing the input in conditions of perfect tracking, i.e., $\tau_0(p, x, \dot{x}, \bar{r}_y, 0)$;
\item a stabilizer is introduced to make the resulting periodic trajectory uniformly orbitally asymptotically stable.
\end{itemize}

\section{Optimal Shaping of the Zero Dynamics}\label{sec:optimization}

Before introducing the optimization strategy, it is significant to investigate the conditions for the solvability of the problem.
In fact, we previously underlined that in general the configuration $\chi = (x, y)$ may leave $\mathcal{N}$.
Intuitively, it is required that, for at least for some values of $p$, there exist bounded trajectories such that the minimal realization \eqref{eq:chi_system} does not incur in singular configurations.
By Assumption \ref{hyp:change_of_coordinates} and due to the input-output linearization strategy, feasibility becomes a feature inherently related with the zero dynamics.
In addition, we need to ensure the generation of (feasible) periodic trajectories.
\par To address both these requirements, we take advantage of the analysis in \cite{shiriaev2006periodic}.
Note that the zero dynamics can be written, for convenience, as:
\begin{equation}\label{eq:zd_x}
\alpha(p, x)\ddot{x} + \beta(p, x)\dot{x}^2 + \gamma(p, x) = 0.
\end{equation}
Indeed, the special structure of system \eqref{eq:zd_x} is obtained from the VHC approach (see \cite[Proposition 2]{shiriaev2005constructive}), considering the constraint $e = y - \bar{r}_y = 0$, while leaving unconstrained the remaining coordinate.

\begin{assumption}\label{hyp:equilibrium}
There exists a compact set $\hat{\mathcal{P}} \subset \mathcal{P}$ such that, for all $p \in \hat{\mathcal{P}}$, there exists a point $x_0(p)$, satisfying $(x_0(p), \bar{r}_y) \in \mathcal{N}$, and such that:
\begin{equation}
\gamma(p, x_0(p)) = 0.
\end{equation}
\end{assumption}

It follows that for all $p \in \hat{\mathcal{P}}$ there exists at least a solution not leaving the set $\mathcal{N}$, preserving the minimal representation \eqref{eq:chi_system}.
Note that in general multiple equilibria may exist.
Exploting Assumption \ref{hyp:equilibrium}, the existence of periodic solutions around $x_0$ is related with the auxiliary linear system:
\begin{equation}\label{eq:aux_system}
\ddot{z} + \underbrace{\left[\frac{\partial}{\partial x}\left(\frac{\gamma(p, x)}{\alpha(p, x)}\right) (p, x_0(p)) \right]}_{\Omega(p)} z = 0.
\end{equation}
Since all sufficient conditions of \cite[Theorem 3]{shiriaev2006periodic} are verified by regularity of $\alpha$, $\beta$, $\gamma$, for all initial conditions $(x(0), \dot{x}(0))$ in a sufficiently small neighborhood of $(x_0(p), 0)$, we obtain that if the above auxiliary system has a center in the origin, i.e.:
\begin{equation}
\Omega(p) > 0,
\end{equation}
then \eqref{eq:zd_x} has a center in $(x_0(p), 0)$.
This result ensures the existence of periodic trajectories for the zero dynamics, locally around $(x_0(p), 0)$.
As a consequence, we introduce the last Assumption for a successful optimization procedure.

\begin{assumption}
There exists $p^* \in \hat{\mathcal{P}}$ such that, for an equilibrium of the form $x_0(p^*)$, it holds $\Omega(p^*) > 0$, i.e., the auxiliary system \eqref{eq:aux_system} has a center in the origin.
\end{assumption}
\par Taking advantage of the previous considerations, we can now define a procedure to select $p$, based on the following intuitive strategy:
\begin{itemize}
\item given the $T_r$-periodic reference $r_x$, consider a time instant $t^*$ satisfying $\dot{r}_x(t^*) = 0$.
This way, we can establish a simple initialization for system \eqref{eq:zd_x};
\item let $s = t - t^*$.
Then, for a given $p \in \mathcal{P}$ such that $\Omega(p) > 0$, compute the trajectory of system \eqref{eq:zd_x} from the initial condition $(r_x(t^*), 0)$.
Indicate such solution, when it exists, with $(\bar{x}(s), \dot{\bar{x}}(s))$, $s \geq 0$;
\item with $(\bar{x}(s), \dot{\bar{x}}(s))$ available, compute $\bar{\tau}(s) = \tau_0(p, \bar{x}(s), \dot{\bar{x}}(s), \bar{r}_y, 0)$;
\item to evaluate the choice of $p$, and thus obtain information to update its selection, consider a cost functional $J(p)$ of the form
\begin{equation}
J = \int_0^{T_r}L(p, r_x(t^* + s), \bar{x}(s), \dot{\bar{x}}(s), \bar{\tau}(s))ds + J_{\text{f}}(p, r_x(t^* + T_r), \bar{x}(T_r), \dot{\bar{x}}(T_r)),
\end{equation}
where $L$ and $J_{\text{f}}$ are non-negative functions to be appropriately selected depending on the design objectives.
\end{itemize}
This approach can be readily converted into an optimization problem, summarized as follows (keeping the dependence of $J$ on $p$ implicit, to better highlight the design features):
\begin{equation}
\begin{split}
\min_{p \in \mathcal{P}} & \int_0^{T_r}L(p, r_x(t^* + s), \bar{x}(s), \dot{\bar{x}}(s), \bar{\tau}(s))ds + J_{\text{f}}(p, r_x(t^* + T_r), \bar{x}(T_r), \dot{\bar{x}}(T_r))\\
& \text{subject to: } \alpha(p, \bar{x}(s))\ddot{\bar{x}}(s) + \beta(p, \bar{x}(s))\dot{\bar{x}}(s)^2 + \gamma(p, \bar{x}(s)) = 0,\\
&\qquad\qquad\qquad\qquad\bar{\tau}(s) = \tau_0(p, \bar{x}(s), \dot{\bar{x}}(s), \bar{r}_y, 0),\\
& \qquad\qquad\qquad  \bar{x}(0) = r_x(t^*), \quad \dot{\bar{x}}(0) = 0, \quad \Omega(p) > 0.
\end{split}
\end{equation}

For the numerical development of this problem, we employed the environment Matlab-Simulink, considering different choices of the decision variables.
In particular, the differential equation \eqref{eq:zd_x} was implemented in Simulink: this way, the optimization routine evaluates the cost function by running simulations with fixed duration $T_r$.
Concerning the algorithm selection, we tested two Matlab global optimization solvers, i.e., a genetic algorithm (\texttt{ga}) and particle swarm (\texttt{particleswarm}).
This choice is motivated by the fact that gradient descent algorithms may easily converge to local minima, which is detrimental to the overall performance.
Two scenarios for optimization are considered:
\begin{enumerate}
\item we optimize the spring stiffness in two passive joints.
The analysis of this case is of particular conceptual interest, since we can investigate the shape of $J(p)$ as a 2-D map;
\item we use as decision variables both the spring stiffness in two passive joints and the shape of two links (without modifying the respective lengths).
Since the material density is supposed constant, this choice allows to adjust (reduce) the weight of the mechanism.
\end{enumerate}
We underline that no damping contribution is present in the mechanism, nor it is introduced to shape its dynamics.
Since we also do not consider gravity, system \eqref{eq:el_dae} depends in the following only on the geometry, the masses, and the optimized elastic components.

\subsection{Optimization of the Spring Stiffness in Two Passive Joints}

\begin{table}[t!]	
	\begin{center}
		\captionsetup{width=0.7\columnwidth}
		\caption{Parameters of the Mechanism Links}\label{tab:link_parameters}
		\addtolength{\tabcolsep}{-4pt}  
		\begin{tabular}{lr | lr | lr}\hline
		        {\scriptsize Length} & {\scriptsize [$\text{m}$]} & {\scriptsize Total mass} & {\scriptsize [$\text{kg}$]} & {\scriptsize Moment of inertia} & {\scriptsize[$\text{kg m}^2$]} \\ 
			\hline
			{\scriptsize $l_1$}  &{\scriptsize $0.080$} & {\scriptsize $m_1$} & {\scriptsize $0.071$} & {\scriptsize $J_1$} & {\scriptsize $0.188\times10^{-3}$}\\ 
			{\scriptsize $l_2$}  &{\scriptsize $0.235$} & {\scriptsize $m_2$} & {\scriptsize $0.195$} & {\scriptsize $J_2$} & {\scriptsize $1.041\times10^{-3}$}\\
			{\scriptsize $l_3$}  &{\scriptsize $0.052$} & {\scriptsize $m_3$ ($\delta_{\text{t}} = 0.02$m)} & {\scriptsize $0.049$} & {\scriptsize $J_3$ ($\delta_\text{t} = 0.02$m)} & {\scriptsize $0.035\times10^{-3}$}\\
			{\scriptsize $l_4$}  &{\scriptsize $0.135$} & {\scriptsize $m_4$ ($\delta_{\text{b}} = 0.02$m)} & {\scriptsize $0.115$} & {\scriptsize $J_4$ ($\delta_\text{b} = 0.02$m)} & {\scriptsize $0.767\times10^{-3}$}\\ 
			\hline
			\end{tabular}
			\addtolength{\tabcolsep}{4pt}
			  
		\end{center}

\end{table}

\begin{figure}[t!]
	\centering
	\psfragscanon
	
	\psfrag{h} [B][B][0.7][0]{$k_{\text{b}}$}
	\psfrag{k} [B][B][0.7][0]{$k_{\text{t}}$}
	\psfrag{m} [B][B][0.7][0]{$l_4$, $\delta_{\text{b}}$}
	\psfrag{n} [B][B][0.7][0]{$l_3$,  $\delta_{\text{t}}$}
	\includegraphics[clip = true, height = 5cm]{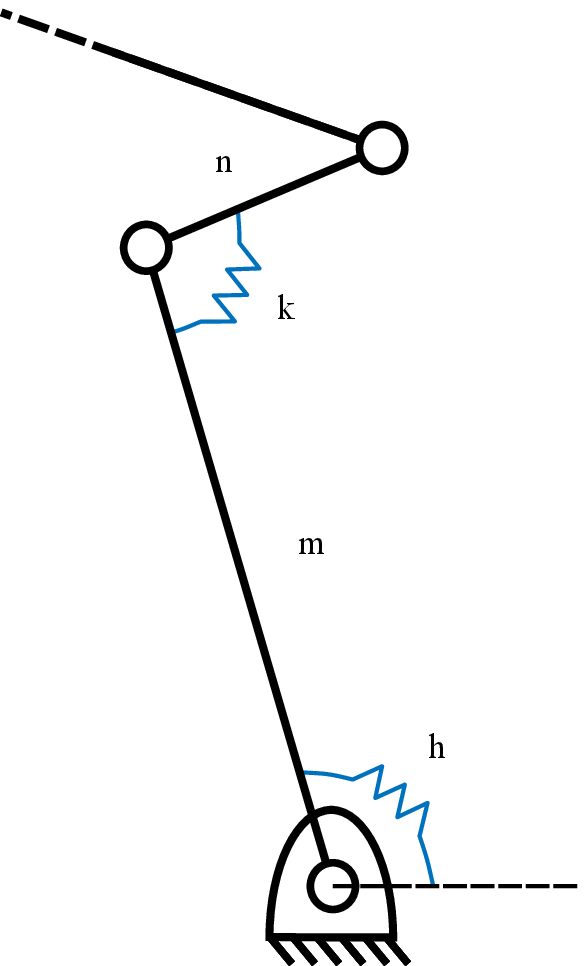}
	\hspace{30pt}
	\psfrag{l} [B][B][0.7][0]{$l$}
	\psfrag{d} [B][B][0.7][0]{$\delta$}
	\includegraphics[clip = true, height = 5cm]{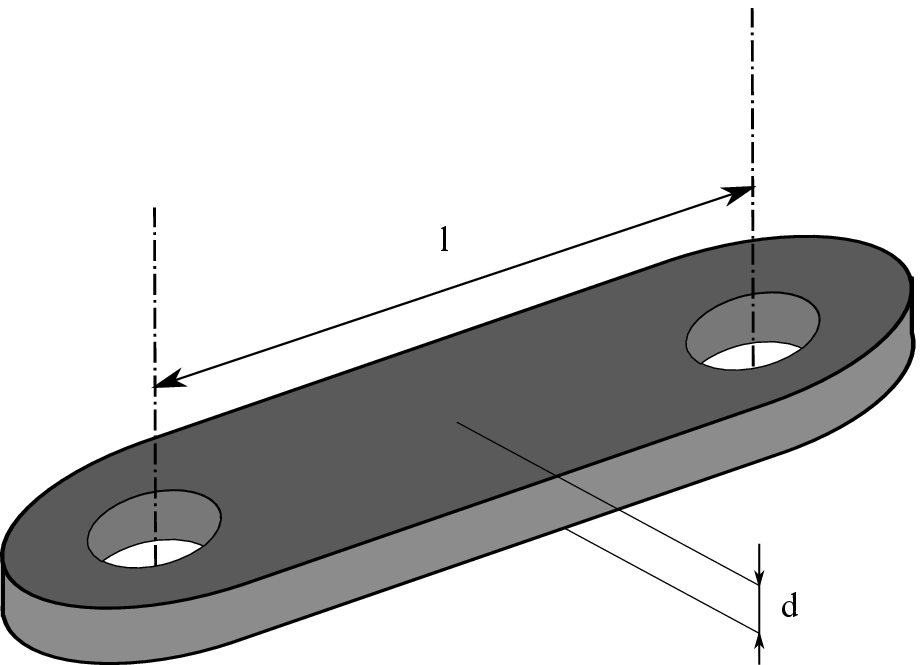}
	
	\caption{\textit{Left:} torsional springs considered for structural optimization. The dashed line connected to $k_{\text{b}}$ indicates the fixed frame of the mechanism. \textit{Right:} Shape of each link, with the depth $\delta$ of the links $3$ and $4$ ($\delta_{\text{t}}$ and $\delta_{\text{b}}$, respectively) used for structural optimization.}
	\label{fig:springs_and_link_structure}
\end{figure}

\begin{figure}[t!]
	\centering
	\psfragscanon
	
	\psfrag{x} [B][B][0.7][0]{$k_{\text{b}}$ [Nm/rad]}
	\psfrag{y} [B][B][0.7][0]{$k_{\text{t}}$ [Nm/rad]}
	\psfrag{t} [B][B][0.7][0]{$J(k_{\text{b}}, k_{\text{t}})$}
	\includegraphics[clip = true, width = 0.32\textwidth]{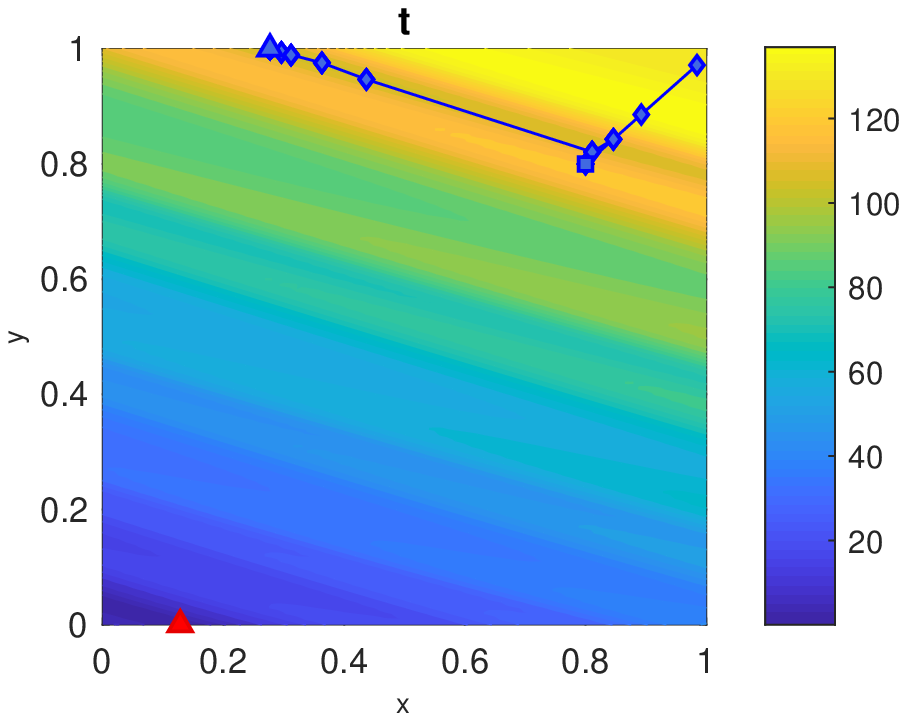}
	\hspace{5pt}
	\psfrag{x} [B][B][0.7][0]{$k_{\text{b}}$ [Nm/rad]}
	\psfrag{y} [B][B][0.7][0]{$k_{\text{t}}$ [Nm/rad]}
	\psfrag{z} [B][B][0.7][0]{$J(k_{\text{b}}, k_{\text{t}})$}
	\includegraphics[clip = true, width = 0.32\textwidth]{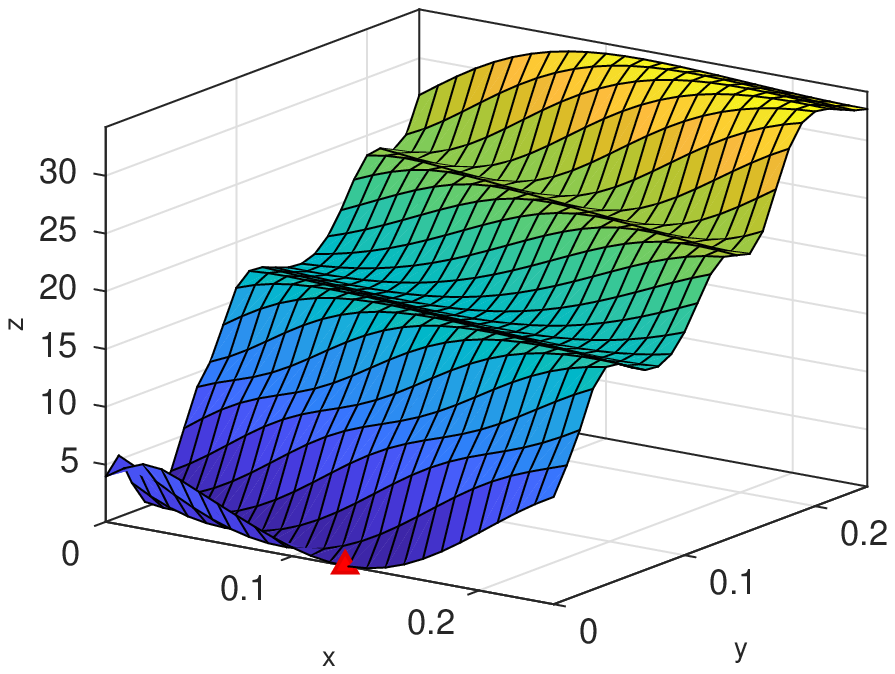}
	
	\caption{Cost function $J$ associated with the optimization of the spring stiffness in two passive joints. \textit{Left:} 2-D plot of the cost function in the overall domain, with the global minimum indicated with a red triangle, and a suboptimal optimization trajectory of the Interior-Point algorithm of \texttt{fmincon} in blue. In particular, the initial guess of this trajectory is given by the blue square, while the final point is denoted with a blue triangle. \textit{Right:} 3-D plot of the cost function in a region surrounding the global minimum.}
	\label{fig:J1}
\end{figure}

In this scenario, we propose to include two springs as shown in Figure \ref{fig:springs_and_link_structure}-Left, with associated torsional stiffness $k_{\text{b}}$, $k_{\text{t}}$.
In the following, we indicate $p = (k_{\text{b}}, k_{\text{t}})$, while $\mathcal{P}$ is given for simplicity by a square box of the form $[0, 1]^2$.
The remaining parameters of the mechanism are summarized in Table \ref{tab:link_parameters}.
In particular, we indicated the moments of inertia with respect to an axis passing through the center of mass (placed in the midpoint of each link).
Finally, the position of the first joint (corresponding to $q_1$) is $(-0.19, 0.15)$m in the $e_x$-$e_y$ frame.
\par For what concerns the design of $J$, we chose a structure analogous to those arising in linear quadratic optimal control.
In particular, for a trajectory of the form $(\bar{x}(s), \dot{\bar{x}}(s))$, let $\epsilon(s) \coloneqq (\bar{x}(s) - r_x(t^* + s), \dot{\bar{x}}(s) - \dot{r}_x(t^* + s))$.
Then, consider:
\begin{equation}\label{eq:J1}
J = \int_0^{T_r}\left[\epsilon(s)^TQ\epsilon(s) + R|\bar{\tau}(s)|^2 \right]ds + \epsilon(T_r)^T S \epsilon(T_r),
\end{equation}
for a positive scalar $R$ and positive-definite matrices $Q$, $S$.
To perform optimization using \texttt{ga}, we included the requirement $\Omega(p) > 0$ as a nonlinear constraint in the solver options.
On the other hand, such a condition was included in \texttt{particleswarm} through a barrier function, designed to grow to $+\infty$ as $\Omega(p) \to 0^+$.
The overall cost function for \texttt{particleswarm} is thus given by $J + B$, where $B$ satisfies:
\begin{equation}
B(p) = -c \log\left(\frac{\Omega(p)}{1 + \Omega(p)}\right),
\end{equation}
with $c > 0$ an arbitrarily small scalar.
\par In the numerical design, we selected $Q = 100 I_2$, $R = 10$, $S = 10 I_2$, leading to the function $J(k_{\text{b}}, k_{\text{t}})$ depicted in Figure \ref{fig:J1}.
We omit the function $B$, implemented in \texttt{particleswarm} with $c = 0.1$, since the only point in $\mathcal{P}$ violating $\Omega(p) > 0$ is given by $k_{\text{b}} = 0$, $ k_{\text{t}} = 0$.
In Figure \ref{fig:J1}, we can appreciate that $J$ is non-convex and includes several local minima, in addition to the global minimum indicated with a red triangle.
This fact motivates the use of the considered solvers, which effectively reach the optimal value $(0.1283, 0) \text{ Nm/rad}$, with a distance between the \texttt{ga} and \texttt{particleswarm} optimal points below $1\times 10^{-5}$.
Notably, other solvers may lead to suboptimal solutions if incorrectly initialized.
This aspect is shown in Figure \ref{fig:J1}-Left, where we indicated a trajectory (in blue) of the Interior-Point algorithm of \texttt{fmincon}, initialized in $(0.8, 0.8) \text{ Nm/rad}$ and converging to $(0.2771, 1) \text{ Nm/rad}$.
\par
\begin{figure}[b!]

	\psfragscanon
	
	\vspace{3pt}
	
	\begin{subfigure}[b]{\textwidth}
	\centering
	\psfrag{x} [B][B][0.7][0]{(a) time [s]}
	\psfrag{y} [B][B][0.7][0]{$\bar{x}$, $r_x$ [m]}
	\includegraphics[clip = true, width = 0.23\textwidth]{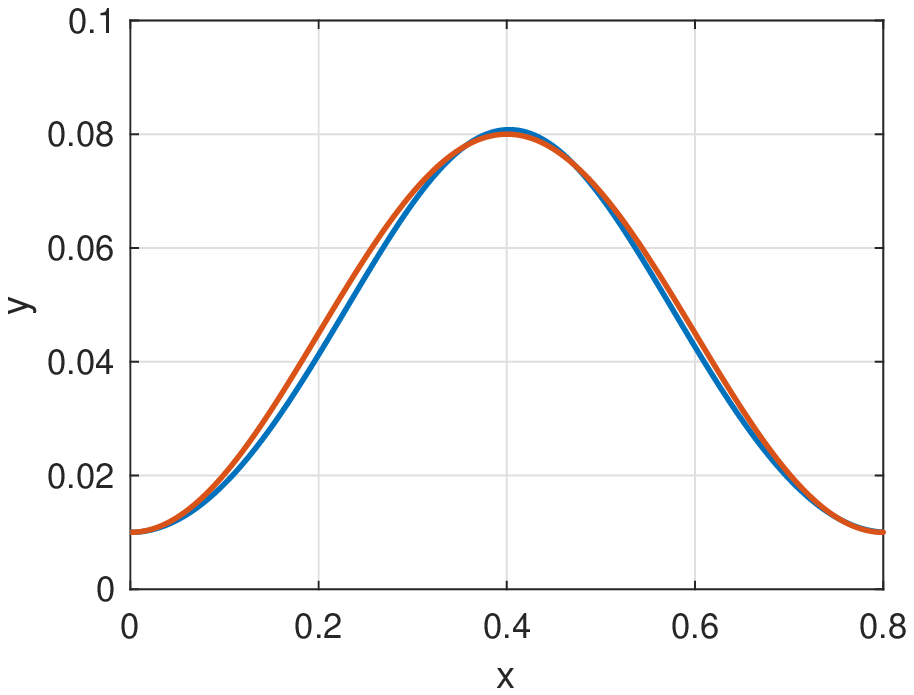}
	\hspace{0.02\textwidth}
	\psfrag{x} [B][B][0.7][0]{(b) time [s]}
	\psfrag{y} [B][B][0.7][0]{$\dot{\bar{x}}$, $\dot{r}_x$ [m/s]}
	\includegraphics[clip = true, width = 0.23\textwidth]{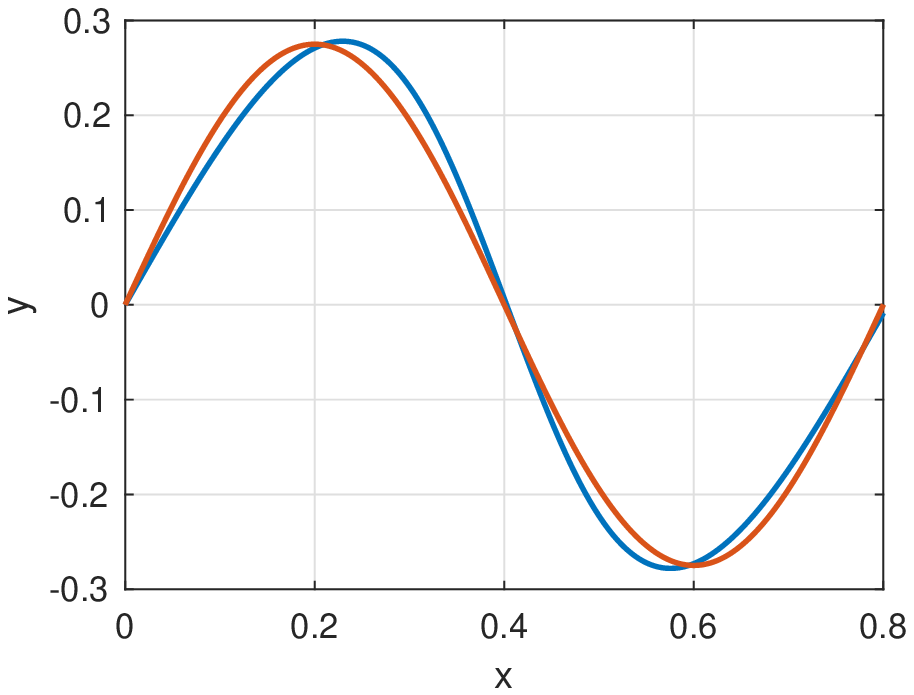}
	\hspace{0.02\textwidth}
	\psfrag{x} [B][B][0.7][0]{(c) $\bar{x}$, $r_x$ [m]}
	\psfrag{y} [B][B][0.7][0]{$\dot{\bar{x}}$, $\dot{r}_x$ [m/s]}
	\includegraphics[clip = true, width = 0.23\textwidth]{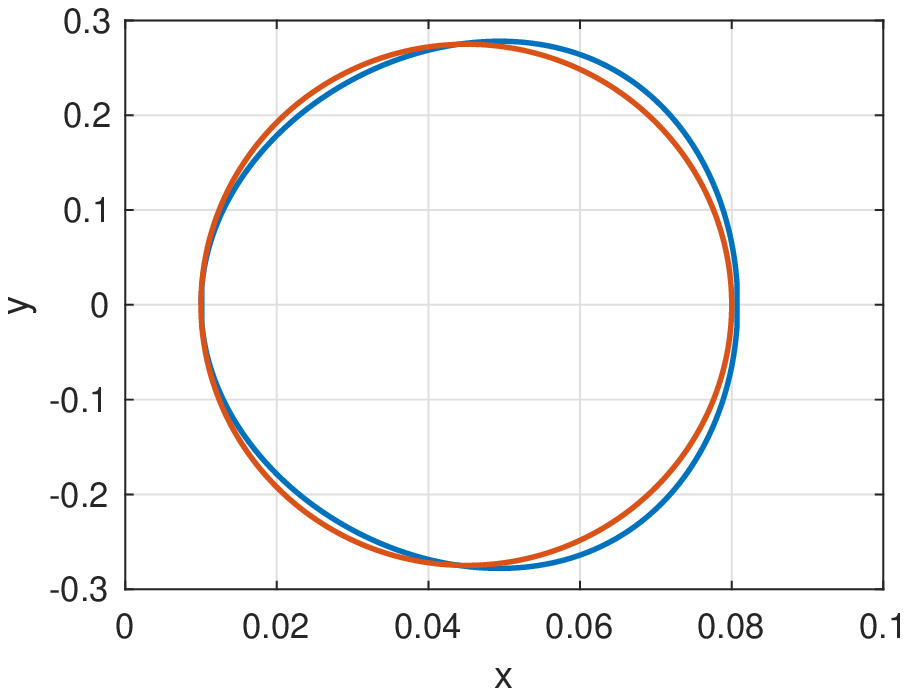}
	\end{subfigure}
	
	\vspace{3pt}
	
	\begin{subfigure}[b]{\textwidth}
	\centering
	\psfrag{x} [B][B][0.7][0]{(d) time [s]}
	\psfrag{y} [B][B][0.7][0]{$\epsilon_1$ [m]}
	\includegraphics[clip = true, width = 0.23\textwidth]{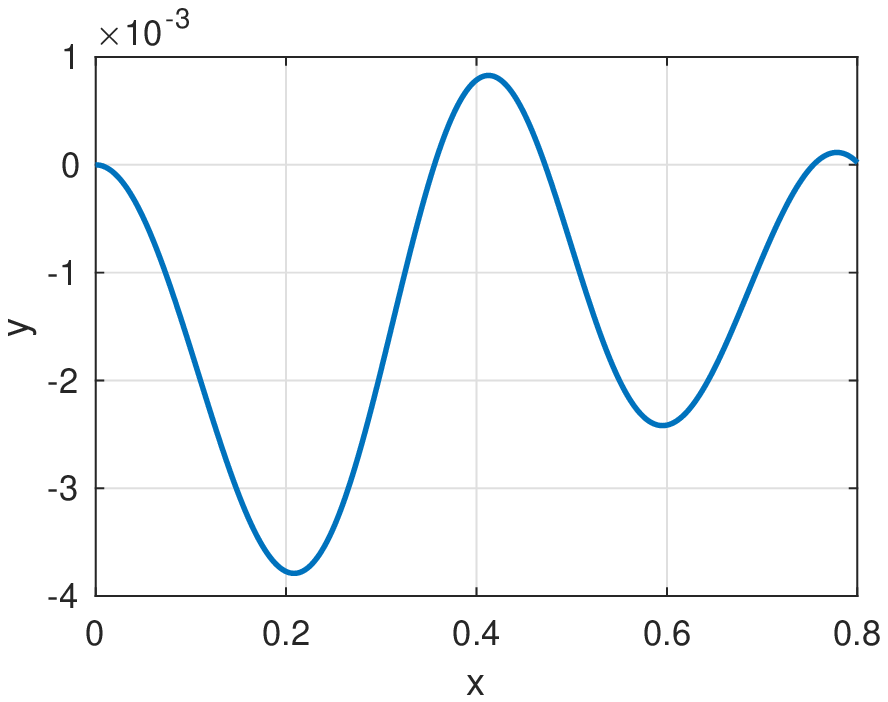}
	\hspace{0.02\textwidth}
	\psfrag{x} [B][B][0.7][0]{(e) time [s]}
	\psfrag{y} [B][B][0.7][0]{$\epsilon_2$ [m/s]}
	\includegraphics[clip = true, width = 0.23\textwidth]{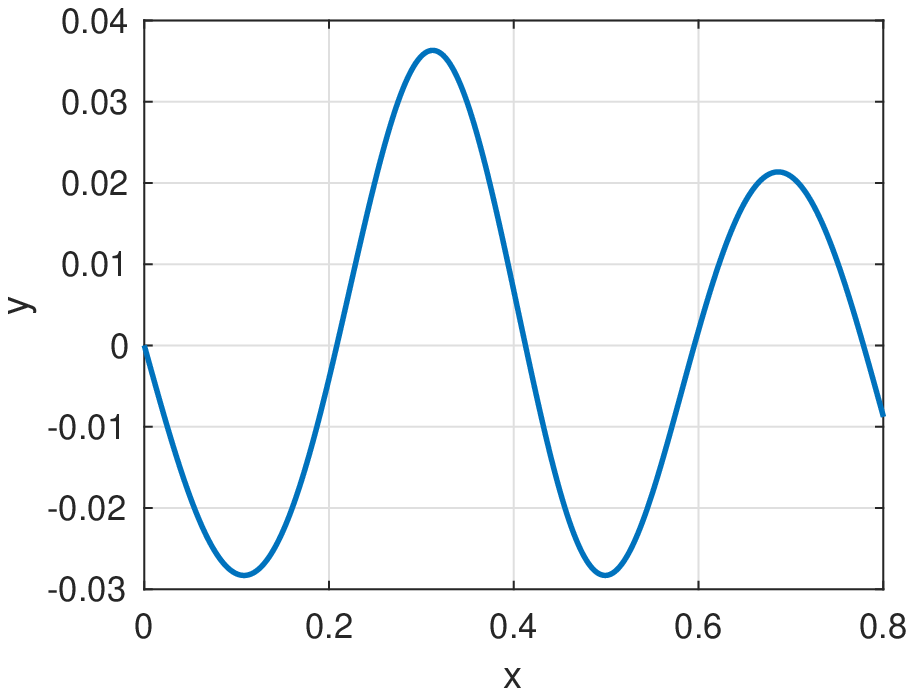}
	\hspace{0.02\textwidth}
	\psfrag{x} [B][B][0.7][0]{(f) time [s]}
	\psfrag{y} [B][B][0.7][0]{$\bar{\tau}$ [Nm]}
	\includegraphics[clip = true, width = 0.23\textwidth]{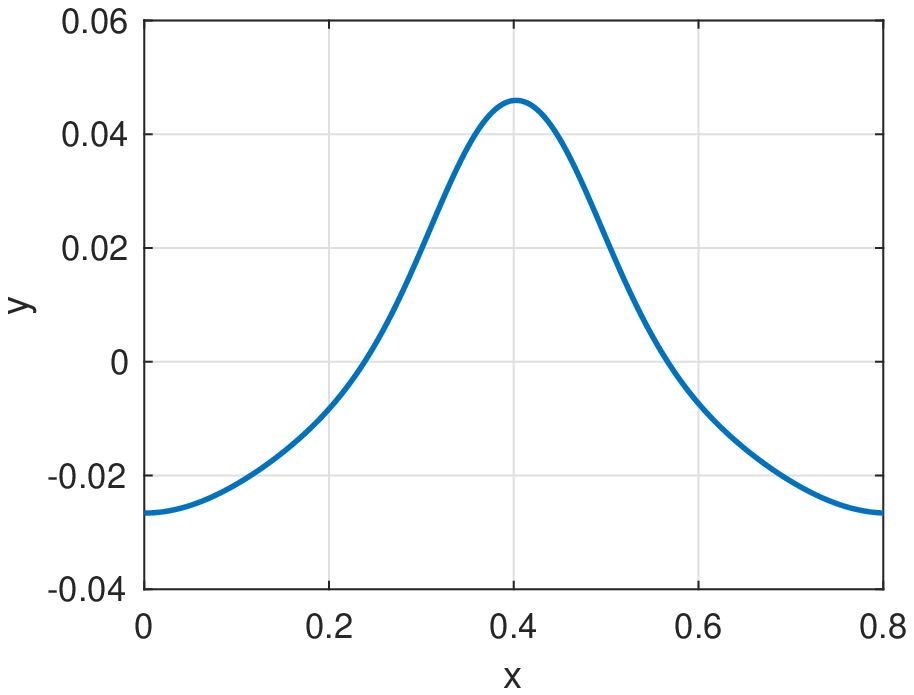}
	\end{subfigure}
		
	\vspace{3pt}
	
	\caption{Results associated with the 2-parameter optimization problem (\texttt{ga} solver). (a): reference $r_x$ (red) and zero dynamics evolution (blue). (b): reference derivative $\dot{r}_x$ (red) and zero dynamics evolution (blue). (c): comparison of the target orbit (red) and the one generated by the zero dynamics (blue). (d): error $\epsilon_1$. (e): error $\epsilon_2$. (f): input $\bar{\tau}$.}
	\label{fig:J1_optimized_trajectories}
\end{figure}

 Regarding the obtained optimal orbit, the results are depicted in Figure \ref{fig:J1_optimized_trajectories}.
There, we can see that the behavior of both $\bar{x}$ (subplot (a)) and $\dot{\bar{x}}$ (subplot (b)) closely matches the target waveforms, as highlighted in the error trajectories (subplots (d)-(e)).
The overall deformation of $(r_x, \dot{r}_x)$ is shown in the phase diagram of Figure \ref{fig:J1_optimized_trajectories}-(c), where the original shape is in red, while the optimized orbit is in blue.
Finally, the behavior of the feedforward torque is given in Figure \ref{fig:J1_optimized_trajectories}-(f).

\subsection{Optimization of the Joint Spring Stiffness and the Link Geometry}

Now we consider a scenario where, in addition to the springs introduced above, we also adjust the shape of the links 3 and 4, while keeping a uniform density ($1000 \text{kg}/\text{m}^3$).
In particular, we propose to adjust the depth in the orthogonal direction w.r.t. the mechanism 2D geometry, as shown in Figure \ref{fig:springs_and_link_structure}-Right.
This way, it is possible to optimize the mass distribution of the system, without modifying the kinematics, entirely preserved from the above scenario.
Note that both masses and moments of inertia scale proportionally with the depth $\delta$ in view of the fixed section and the uniform density.
We remark that other strategies could be adopted, e.g., adding weights in specific positions of the link (see \cite{seifried2014dynamics}).
\par
\begin{figure}[t!]

	\psfragscanon
	
	\vspace{3pt}
	
	\begin{subfigure}[b]{\textwidth}
	\centering
	\psfrag{x} [B][B][0.7][0]{(a) time [s]}
	\psfrag{y} [B][B][0.7][0]{$\bar{x}$, $r_x$ [m]}
	\includegraphics[clip = true, width = 0.23\textwidth]{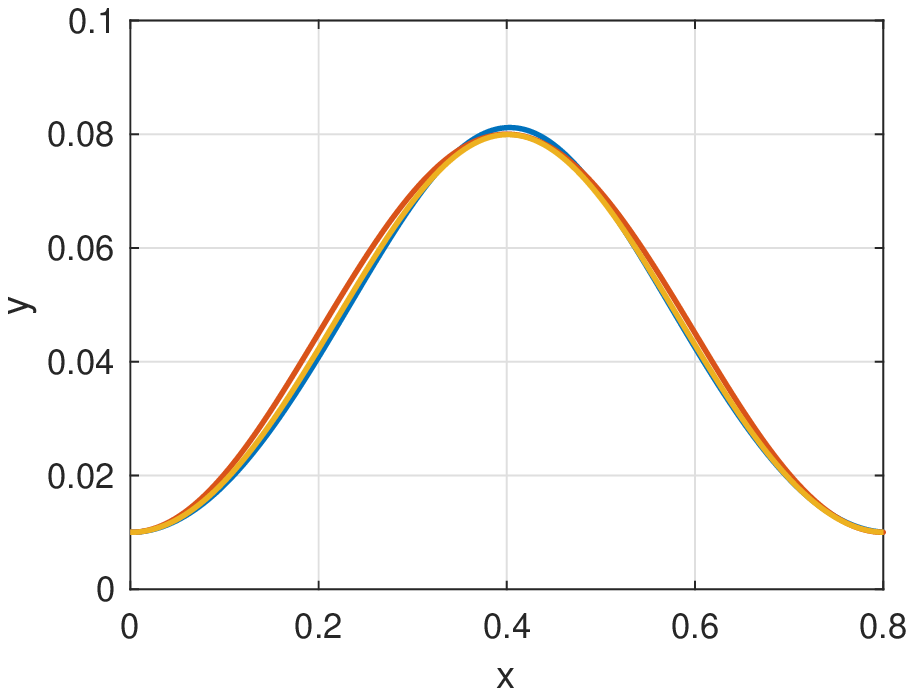}
	\hspace{0.02\textwidth}
	\psfrag{x} [B][B][0.7][0]{(b) time [s]}
	\psfrag{y} [B][B][0.7][0]{$\dot{\bar{x}}$, $\dot{r}_x$ [m/s]}
	\includegraphics[clip = true, width = 0.23\textwidth]{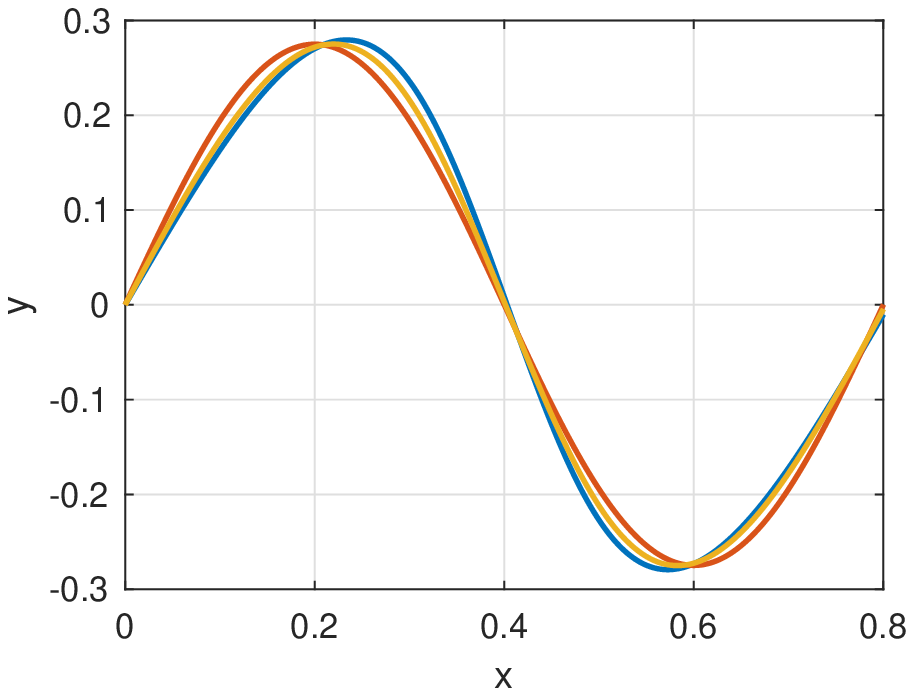}
	\hspace{0.02\textwidth}
	\psfrag{x} [B][B][0.7][0]{(c) $\bar{x}$, $r_x$ [m]}
	\psfrag{y} [B][B][0.7][0]{$\dot{\bar{x}}$, $\dot{r}_x$ [m/s]}
	\includegraphics[clip = true, width = 0.23\textwidth]{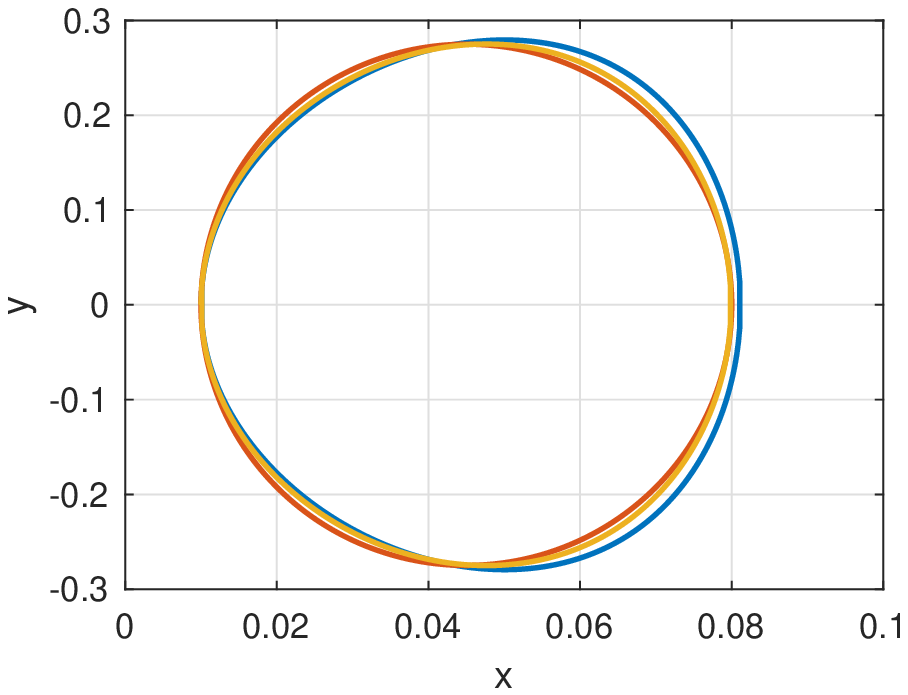}
	\end{subfigure}
	
	\vspace{3pt}
	
	\begin{subfigure}[b]{\textwidth}
	\centering
	\psfrag{x} [B][B][0.7][0]{(d) time [s]}
	\psfrag{y} [B][B][0.7][0]{$\epsilon_1$ [m]}
	\includegraphics[clip = true, width = 0.23\textwidth]{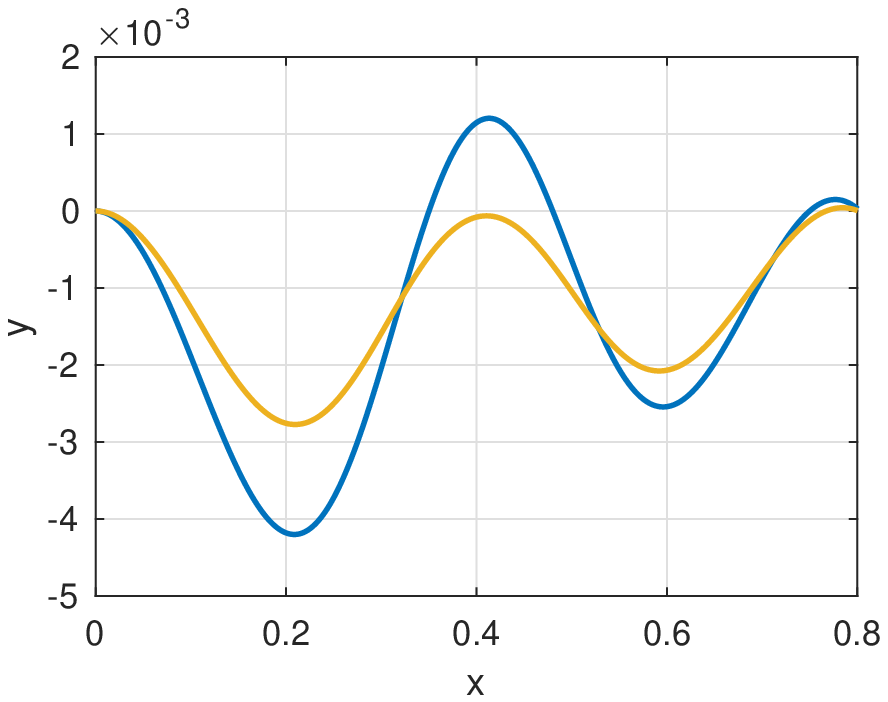}
	\hspace{0.02\textwidth}
	\psfrag{x} [B][B][0.7][0]{(e) time [s]}
	\psfrag{y} [B][B][0.7][0]{$\epsilon_2$ [m/s]}
	\includegraphics[clip = true, width = 0.23\textwidth]{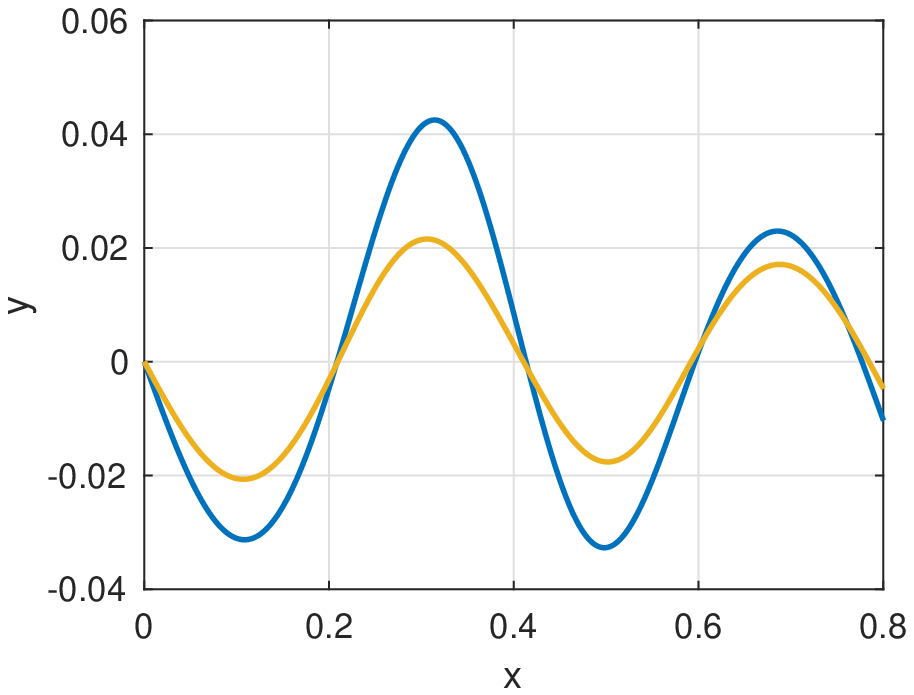}
	\hspace{0.02\textwidth}
	\psfrag{x} [B][B][0.7][0]{(f) time [s]}
	\psfrag{y} [B][B][0.7][0]{$\bar{\tau}$ [Nm]}
	\includegraphics[clip = true, width = 0.23\textwidth]{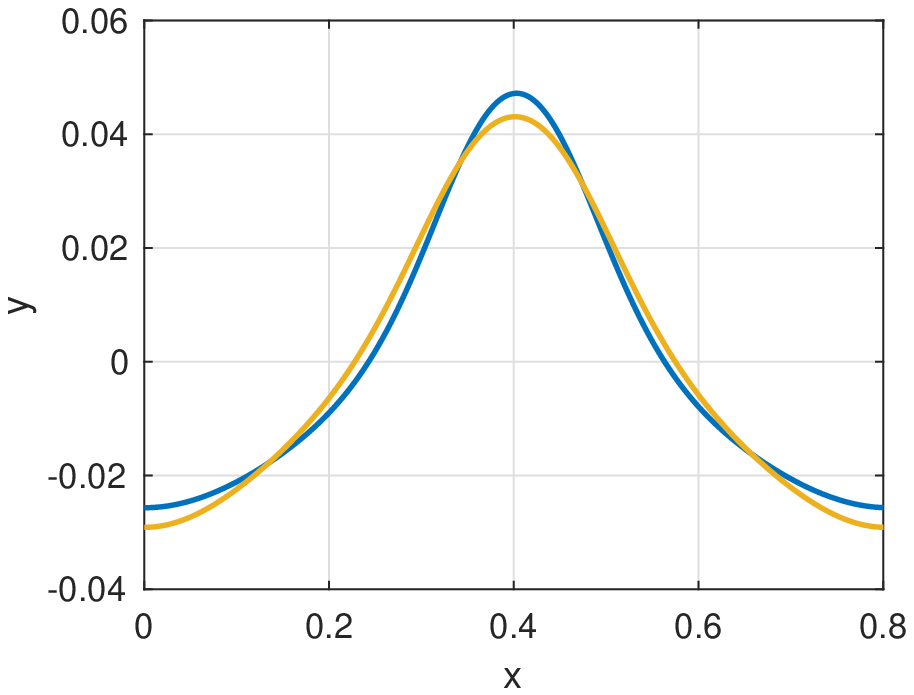}
	\end{subfigure}
		
	\vspace{3pt}
	
	\caption{Results associated with the 4-parameter optimization problem (\texttt{ga} solver). The blue plots are obtained from \eqref{eq:J2}, whereas the yellow plots derive from \eqref{eq:J1}. (a): reference $r_x$ (red) and zero dynamics evolution (blue). (b): reference derivative $\dot{r}_x$ (red) and zero dynamics evolution (blue). (c): comparison of the target orbit (red) and the one generated by the zero dynamics (blue). (d): error $\epsilon_1$. (e): error $\epsilon_2$. (f): input $\bar{\tau}$.}
	\label{fig:J2_optimized_trajectories}
\end{figure}

 Let $p = (k_{\text{b}}, k_{\text{t}}, \delta_{\text{b}}, \delta_{\text{t}})$, where $\delta_{\text{b}}$ and $\delta_{\text{t}}$ are the widths relative to links 4 and 3, respectively.
The set $\mathcal{P}$ that we consider for the decision variables is given by a box of the form $[0, 1]^2\times[0.01, 0.03]^2$.
In particular, $\delta_{\text{b}} = \delta_{\text{t}} = 0.02$m corresponds to the mass distribution of the above 2-parameter optimization.
This way, it is easy to notice that the previous parameter set is a proper subset of the 4-parameter version.
It follows that selecting \eqref{eq:J1} as cost function, with the same weights, yields an optimal cost smaller or equal than the previous one.
Notably, the corresponding optimal solution is found in $(0.173 \text{Nm}/\text{rad}, 0 \text{Nm}/\text{rad}, 0.03 \text{m}, 0.03 \text{m})$ (by both \texttt{ga} and \texttt{particleswarm}), i.e., the constraints for $\delta_{\text{b}}$ and $\delta_{\text{t}}$ are active.
The performance of this solution is validated in Figure \ref{fig:J2_optimized_trajectories}, where in yellow we indicated the corresponding simulated trajectories.
\par Unfortunately, this result is also undesirable because it involves a significant increase of the mass.
Therefore, we introduce as in \cite{bastos2019synergistic} an additional term in the cost function, aiming to penalize the link weight.
In particular, let $\delta = (\delta_{\text{b}}, \delta_{\text{t}}) - (0.01, 0.01)$, then consider in place of \eqref{eq:J1} the following cost function:
\begin{equation}\label{eq:J2}
J = \int_0^{T_r}\left[\epsilon(s)^TQ\epsilon(s) + R|\bar{\tau}(s)|^2 \right]ds + \epsilon(T_r)^T S \epsilon(T_r) + \delta^TL\delta,
\end{equation}
with $L$ a positive definite matrix.
For the numerical results, we selected the gains $Q = 100 I_2$, $R = 10$, $S = 10 I_2$, $L = 200 I_2$, which yield (for both \texttt{ga} and \texttt{particleswarm}, with the same implementation details shown above) the optimal parameters $(0.116 \text{Nm}/\text{rad}, 0 \text{Nm}/\text{rad}, 0.0183 \text{m}, 0.016 \text{m})$.
This way, both the masses and the non-zero stiffness are now reduced (cf. $k_\text{b} = 0.128\text{Nm}/\text{rad}$ from the 2-parameter design).
In particular, the resulting masses are given by $m_3 = 0.039$kg ($20\%$ decrease) and $m_4 = 0.105$kg ($8.7\%$ decrease).
\par The trajectories corresponding to this optimal solution are shown in blue in Figure \ref{fig:J2_optimized_trajectories}, where it is evident how the performance is slightly worse with respect to the maximized mass case.
However, the cost \eqref{eq:J2} allows to achieve a desirable trade-off between mass minimization and trajectory accuracy, which can be suitably adjusted depending on the designer requirements.

\section{Exponential Orbital Stabilization}\label{sec:control}
In this section, we present some numerical results showing that the optimized trajectories can be orbitally stabilized, thus completing the integrated design of Problem \ref{general_problem}.
In particular, \cite[Theorem 3]{shiriaev2005constructive} is exploited to yield exponential orbital stability of trajectories of the form $(\bar{x}(\cdot), \dot{\bar{x}}(\cdot), \bar{r}_y, 0)$.
We report some computational steps involved in the definition of the controller.
\par Recall $e = y - \bar{r}_y$, $\dot{e} = \dot{y}$, then system \eqref{eq:partially_lin_system} can be factorized (keeping the dependence on $\bar{r}_y$ implicit) as:
\begin{equation}
\begin{split}
\ddot x &= - \frac{\beta(p, x)}{\alpha(p, x)} \dot x^2 - \frac{\gamma(p, x)}{\alpha(p, x)} + \frac{\tilde g_{e}(p, x, \dot x, e, \dot{e})}{\alpha(p, x)} e + \frac{\tilde g_{\dot{e}}(p, x, \dot x, e, \dot{e})}{\alpha(p, x)} \dot{e} + \frac{\tilde g_{v}(p, x, e)}{\alpha(p, x)}v \\
\ddot y &= v.
\end{split}
\end{equation}
For a given optimal value $\bar{p}$, indicate with $(\bar{x}(t), \dot{\bar{x}}(t))$ the resulting $T_q$-periodic orbit of the zero dynamics.
Let $\zeta = (\mathcal{I}, e, \dot{e})$, with $\mathcal{I}$ given by \cite{shiriaev2005constructive}:
\begin{equation}
\begin{split}
\mathcal{I} \left(x, \dot{x}, \bar{x}(0), \dot{\bar{x}}(0) \right) &= \dot{x}^2 - \Psi\left(\bar{x}(0), x \right) \left[ \dot{\bar{x}}(0)^2 - 2 \int_{\bar{x}(0)}^{x} \Psi\left(z, \bar{x}(0)\right) \frac{\gamma(\bar{p}, z)}{\alpha(\bar{p}, z)} d z \right]\\
\Psi(a, b) &= \exp \left( -2 \int_{a}^{b} \frac{\beta(\bar{p}, z)}{\alpha(\bar{p}, z)}  dz \right).
\end{split}
\end{equation}
Then, a periodic system is obtained by linearizing the dynamics of $\zeta$ around the origin:
\begin{equation}\label{eq:lin_periodic_system}
\dot \zeta = A(t) \zeta + B(t) v,
\end{equation}
where $A(\cdot)$ and $B(\cdot)$ are $T_q$-periodic matrices computed as
\begin{equation}
A(t) =
\begin{pmatrix}
-\frac{2\dot{\bar{x}}(t)\beta(\bar{p}, \bar{x}(t))}{\alpha(\bar{p}, \bar{x}(t))} & \frac{2\dot{\bar{x}}(t)\tilde g_{e}(\bar{p}, \bar{x}(t), \dot{\bar{x}}(t), 0, 0)}{\alpha(\bar{p}, \bar{x}(t))} & \frac{2\dot{\bar{x}}(t)\tilde g_{\dot{e}}(\bar{p}, \bar{x}(t), \dot{\bar{x}}(t), 0, 0)}{\alpha(\bar{p}, \bar{x}(t))}\\
0 & 0 & 1 \\
0 & 0 & 0
\end{pmatrix}, \qquad
B(t) =
\begin{pmatrix}
\frac{2\dot{\bar{x}}(t)\tilde g_{v}(\bar{p}, \bar{x}(t), 0)}{\alpha(\bar{p}, \bar{x}(t))} \\
0 \\
1
\end{pmatrix}.
\end{equation}
This way, assuming that the pair $(A(t), B(t))$ is completely controllable in the interval $[0, T_q]$, we can assign
\begin{equation}\label{eq:v_law}
v(\bar{p}, x, \dot{x}, \zeta, t) = - \frac{1}{R_{\text{c}}}\begin{pmatrix}\frac{2\dot{x}\tilde g_{v}(\bar{p}, x, e)}{\alpha(\bar{p}, x)} & 0 & 1 \end{pmatrix} P(t) \zeta,
\end{equation}
with $P(t)$ the (positive-definite) periodic solution of the differential Riccati equation:
\begin{equation}
\dot P(t) + A(t)^T P(t) + P(t) A(t) + Q_{\text{c}} = P(t) B(t) R_{\text{c}}^{-1} B(t)^T P(t)
\end{equation}
where $Q_{\text{c}}$ is a positive-definite matrix that penalizes the transient of $\zeta$, while $R_{\text{c}}$ is a positive scalar penalizing the input $v$.
For design, we selected $Q_{\text{c}} = \diag(100, 500, 100)$, $R_{\text{c}} = 1$.
The overall controller is thus given by \eqref{eq:i_o_lin}-\eqref{eq:v_law}.
\par
\begin{figure}[h!]

	\psfragscanon
	
	\vspace{3pt}
	
	\begin{subfigure}[b]{\textwidth}
	\centering
	\psfrag{x} [B][B][0.7][0]{(a) time [s]}
	\psfrag{y} [B][B][0.7][0]{$y$, $\bar{r}_y$ [m]}
	\includegraphics[clip = true, width = 0.23\textwidth]{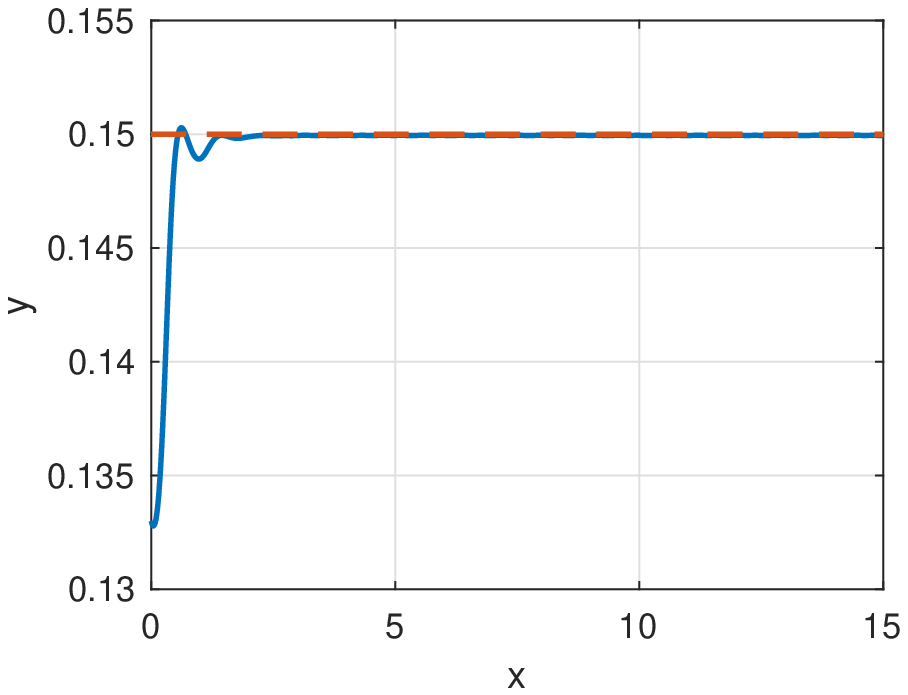}
	\hspace{0.001\textwidth}
	\psfrag{x} [B][B][0.7][0]{(b) time [s]}
	\psfrag{y} [B][B][0.7][0]{$\dot{y}$ [m/s]}
	\includegraphics[clip = true, width = 0.23\textwidth]{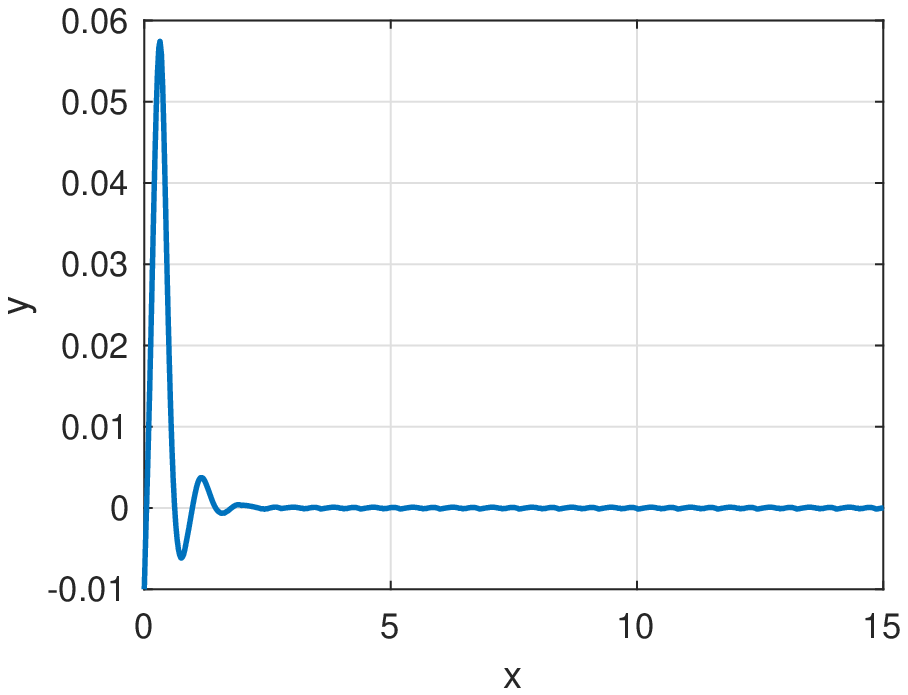}
	\hspace{0.001\textwidth}
	\psfrag{x} [B][B][0.7][0]{(c) $x$, $\bar{x}$ [m]}
	\psfrag{y} [B][B][0.7][0]{$\dot{x}$, $\dot{\bar{x}}$ [m/s]}
	\includegraphics[clip = true, width = 0.23\textwidth]{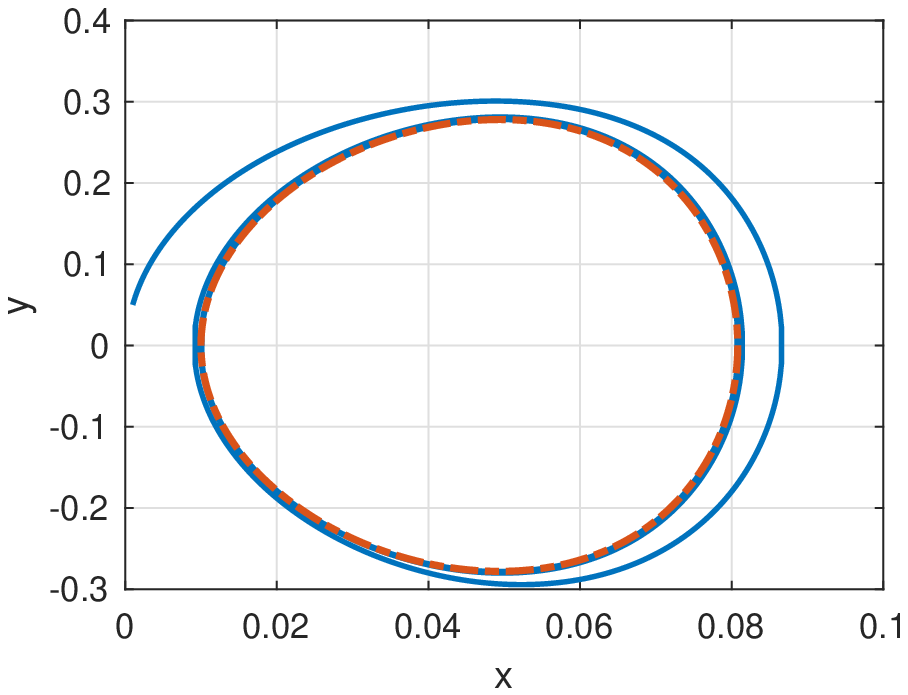}
	\hspace{0.001\textwidth}
	\psfrag{x} [B][B][0.7][0]{(d) time [s]}
	\psfrag{y} [B][B][0.7][0]{$x$ [m]}
	\includegraphics[clip = true, width = 0.23\textwidth]{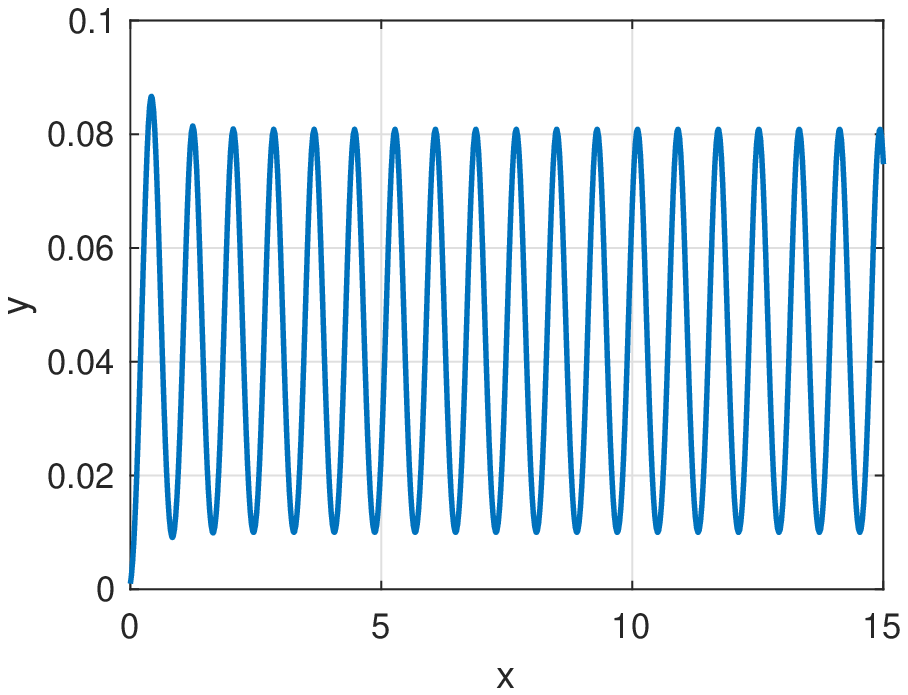}
	\end{subfigure}
	
	\vspace{3pt}
	
	\begin{subfigure}[b]{\textwidth}
	\centering
	\psfrag{x} [B][B][0.7][0]{(e) time [s]}
	\psfrag{y} [B][B][0.7][0]{$e$ [m]}
	\includegraphics[clip = true, width = 0.23\textwidth]{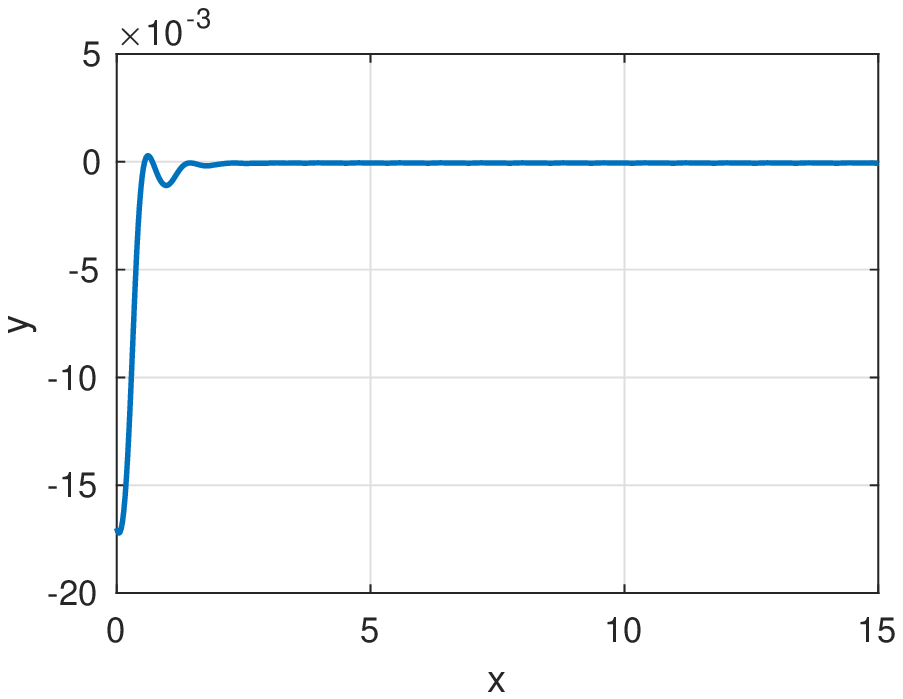}
	\hspace{0.001\textwidth}
	\psfrag{x} [B][B][0.7][0]{(f) time [s]}
	\psfrag{y} [B][B][0.7][0]{$u$ [Nm]}
	\includegraphics[clip = true, width = 0.23\textwidth]{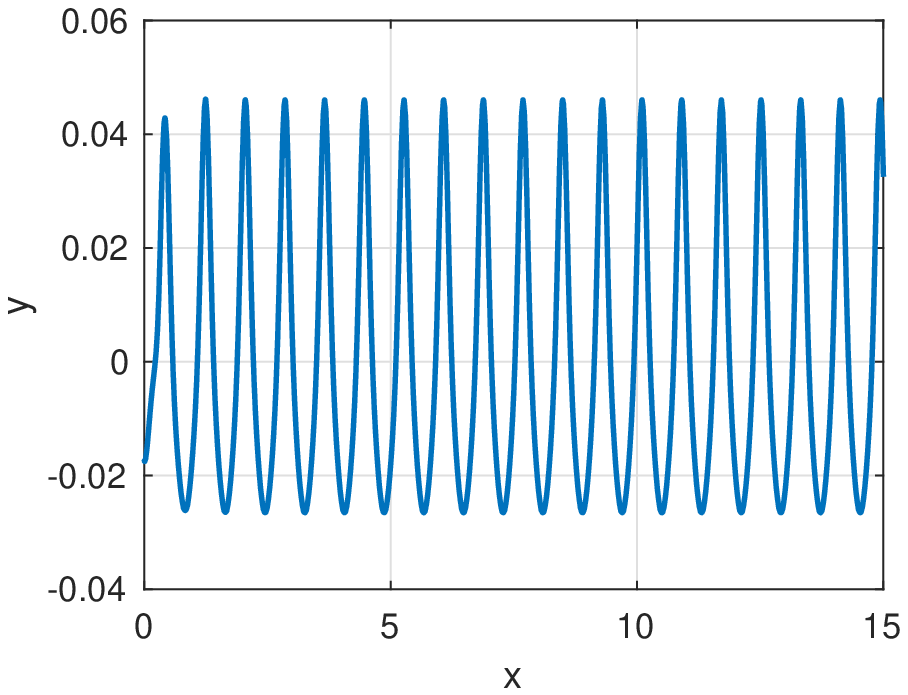}
	\hspace{0.001\textwidth}
	\psfrag{x} [B][B][0.7][0]{(g) time [s]}
	\psfrag{y} [B][B][0.7][0]{$d((x, \dot{x}), \Gamma_x)$}
	\includegraphics[clip = true, width = 0.23\textwidth]{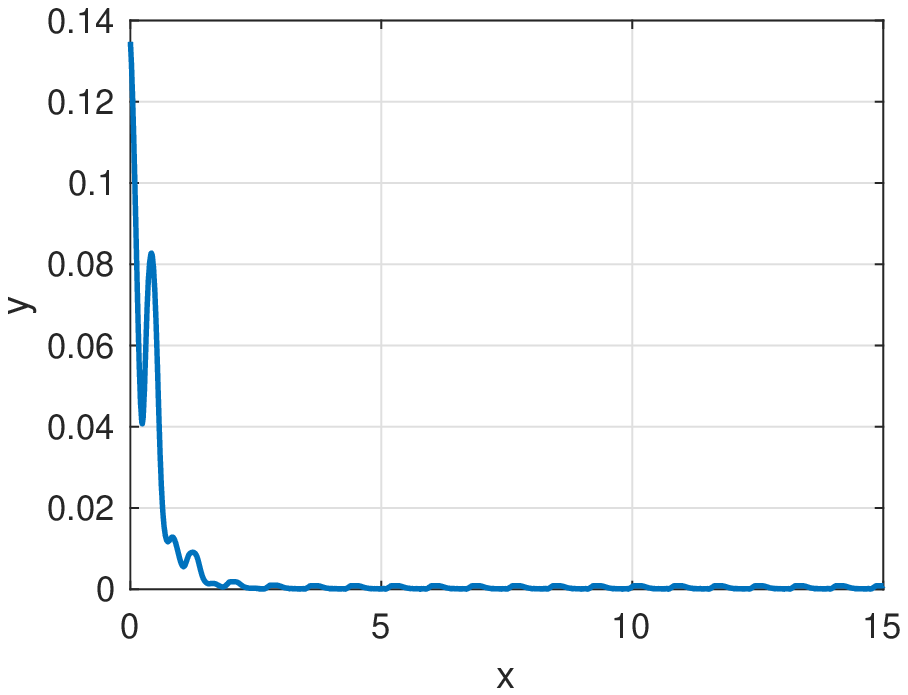}
	\hspace{0.001\textwidth}
	\psfrag{x} [B][B][0.7][0]{(h) time [s]}
	\psfrag{y} [B][B][0.7][0]{$q$ [rad]}
	\psfrag{a} [B][B][0.5][0]{ $q_1$}
	\psfrag{b} [B][B][0.5][0]{ $q_2$}
	\psfrag{c} [B][B][0.5][0]{ $q_3$}
	\psfrag{d} [B][B][0.5][0]{ $q_4$}
	\includegraphics[clip = true, width = 0.23\textwidth]{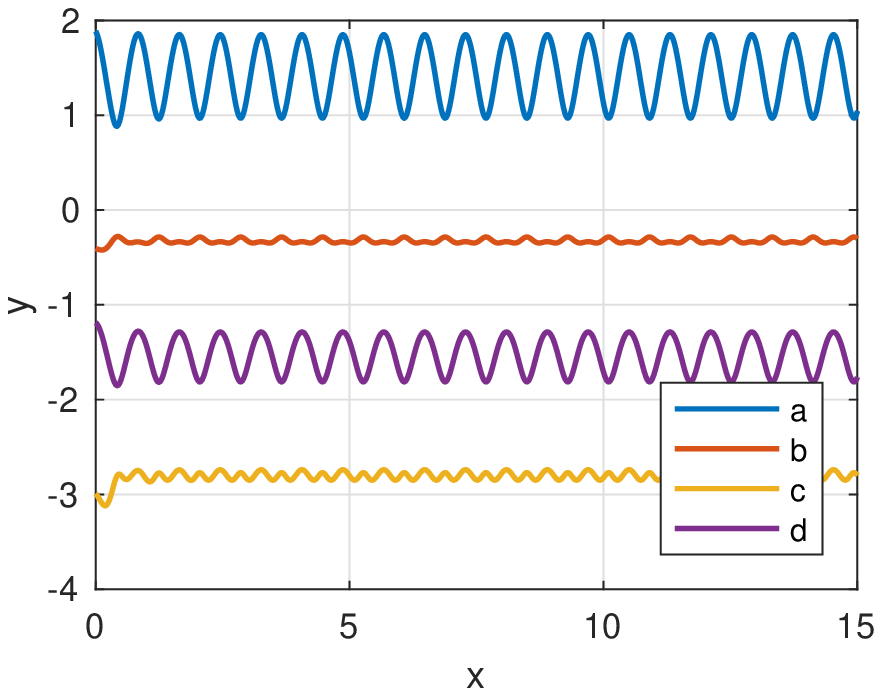}
	\end{subfigure}
		
	\vspace{3pt}
	
	\caption{Orbital stabilization results, with $k_{\text{b}} = 0.1283 \text{Nm}/\text{rad}$, $k_{\text{t}} = 0 \text{Nm}/\text{rad}$ (obtained from the 2-parameter optimization), and $\delta_{\text{b}} = \delta_{\text{t}} = 0.02\text{m}$. (a): $y$ (blue) and reference $\bar{r}_y$ (red). (b): $\dot{y}$. (c): evolution of $(x, \dot{x})$ (blue) and $(\bar{x}, \dot{\bar{x}})$ (red). (d): periodic behavior of $x$. (e): tracking error $e$. (f): control input $u$. (e): normalized distance between $(x, \dot{x})$ and the orbit $\Gamma_x$. (f): behavior of the joint angles.}
	\label{fig:control_simulation1}
\end{figure}

\begin{figure}[h!]

	\psfragscanon
	
	\vspace{3pt}
	
	\begin{subfigure}[b]{\textwidth}
	\centering
	\psfrag{x} [B][B][0.7][0]{(a) time [s]}
	\psfrag{y} [B][B][0.7][0]{$y$, $\bar{r}_y$ [m]}
	\includegraphics[clip = true, width = 0.23\textwidth]{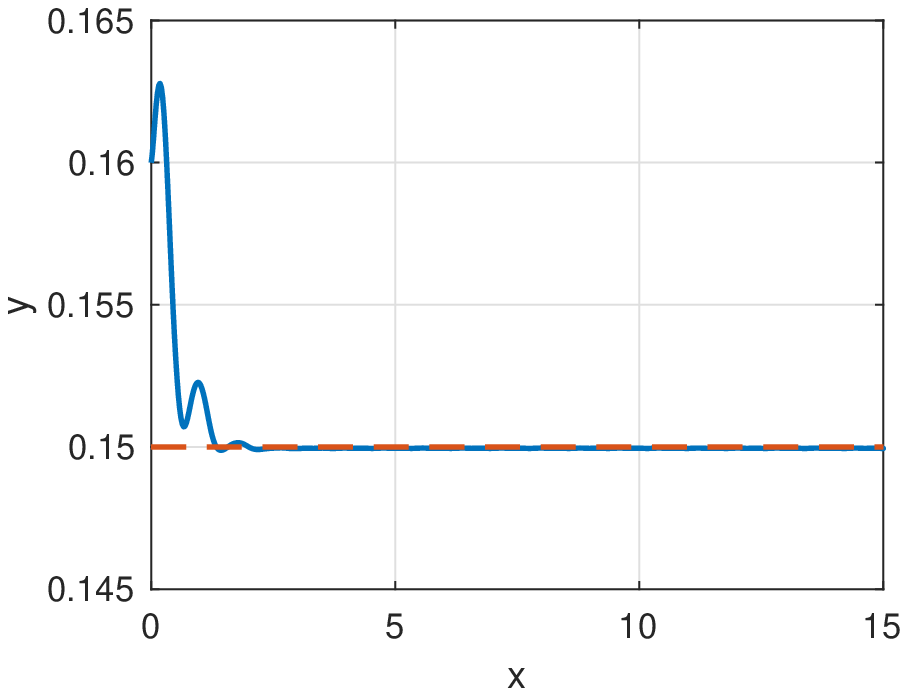}
	\hspace{0.001\textwidth}
	\psfrag{x} [B][B][0.7][0]{(b) time [s]}
	\psfrag{y} [B][B][0.7][0]{$\dot{y}$ [m/s]}
	\includegraphics[clip = true, width = 0.23\textwidth]{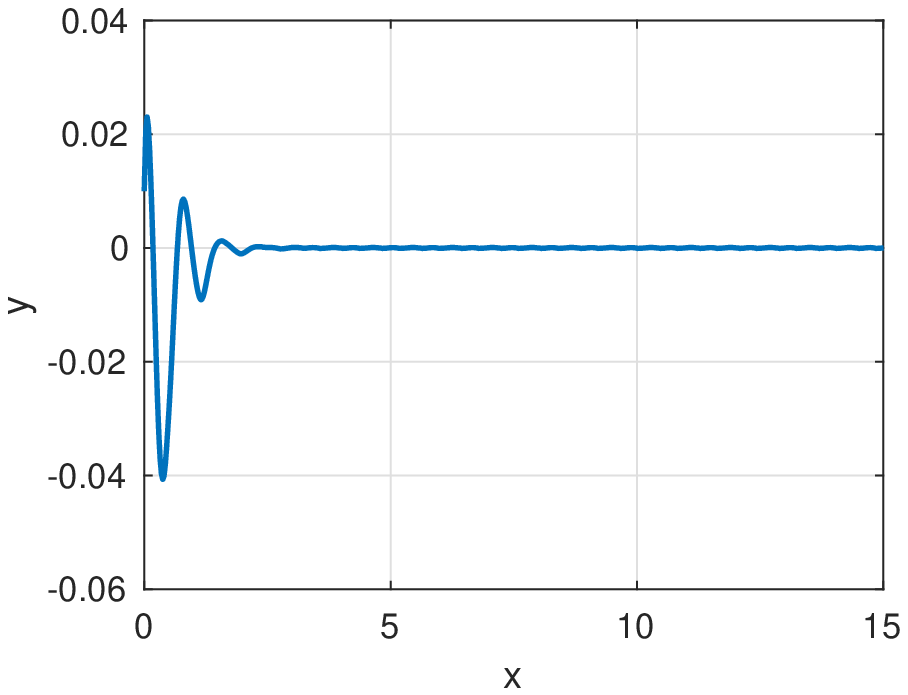}
	\hspace{0.001\textwidth}
	\psfrag{x} [B][B][0.7][0]{(c) $x$, $\bar{x}$ [m]}
	\psfrag{y} [B][B][0.7][0]{$\dot{x}$, $\dot{\bar{x}}$ [m/s]}
	\includegraphics[clip = true, width = 0.23\textwidth]{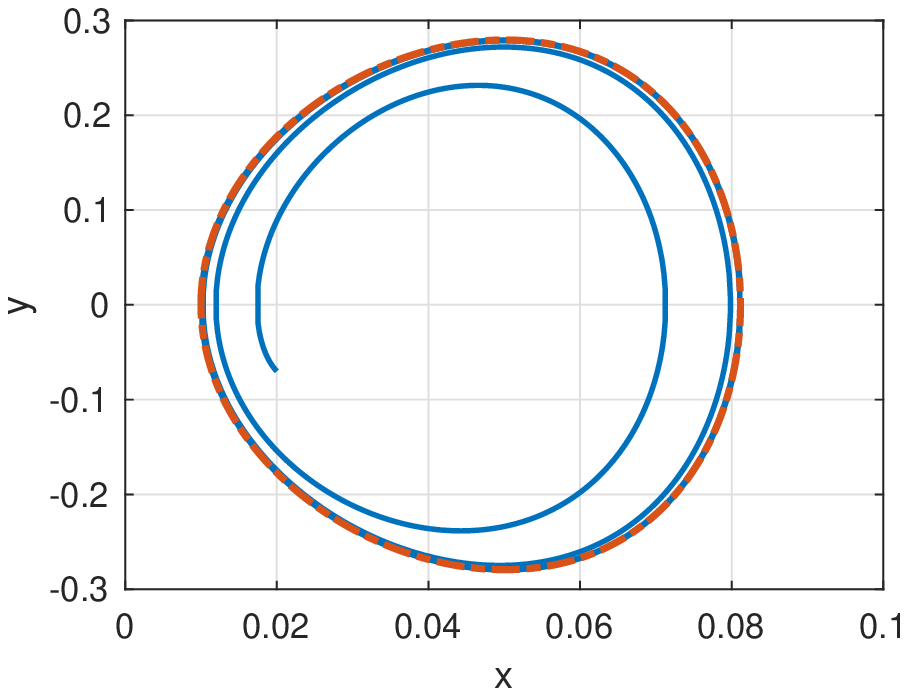}
	\hspace{0.001\textwidth}
	\psfrag{x} [B][B][0.7][0]{(d) time [s]}
	\psfrag{y} [B][B][0.7][0]{$x$ [m]}
	\includegraphics[clip = true, width = 0.23\textwidth]{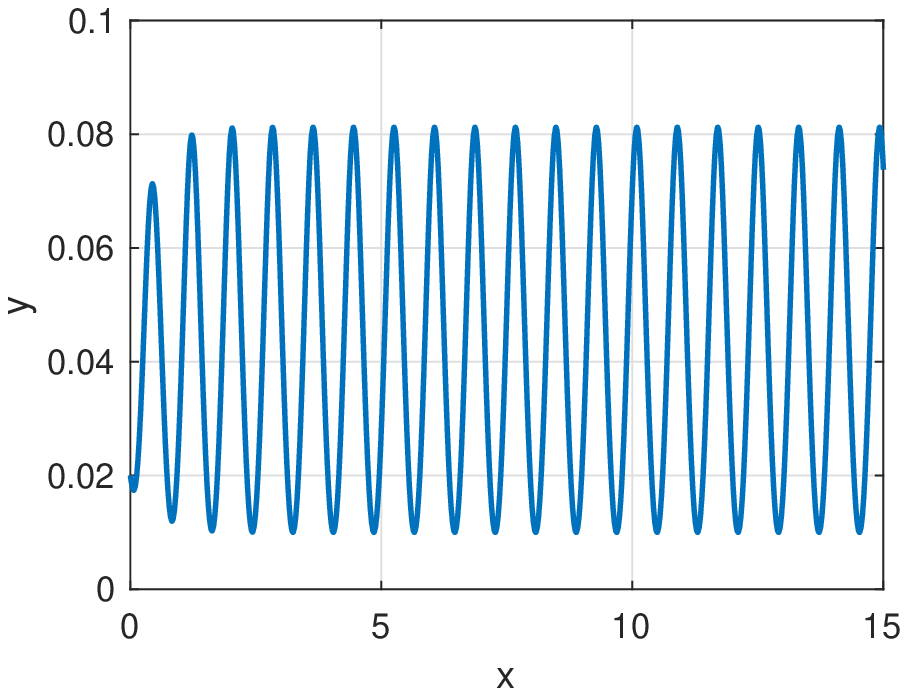}
	\end{subfigure}
	
	\vspace{3pt}
	
	\begin{subfigure}[b]{\textwidth}
	\centering
	\psfrag{x} [B][B][0.7][0]{(e) time [s]}
	\psfrag{y} [B][B][0.7][0]{$e$ [m]}
	\includegraphics[clip = true, width = 0.23\textwidth]{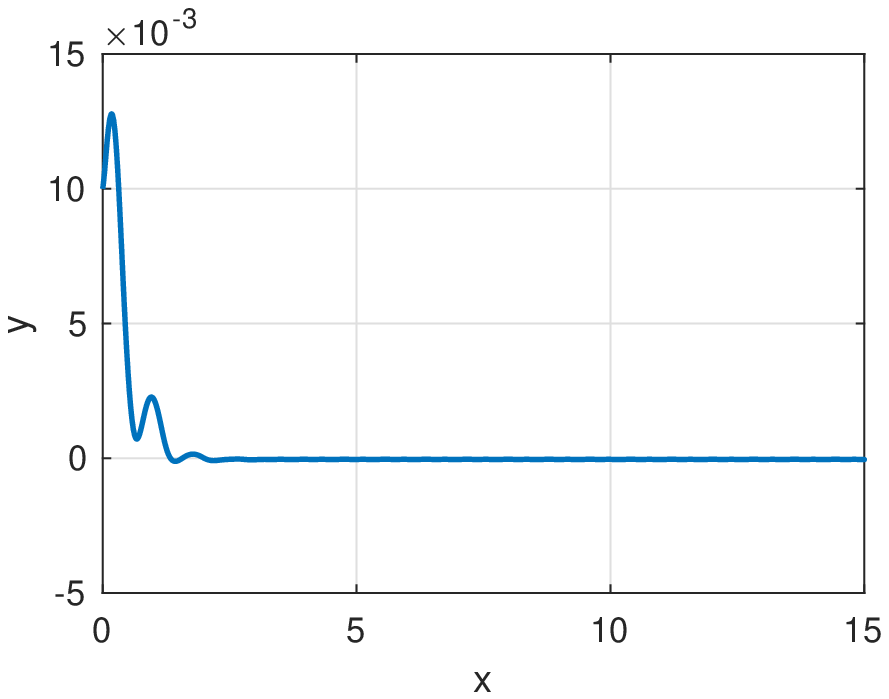}
	\hspace{0.001\textwidth}
	\psfrag{x} [B][B][0.7][0]{(f) time [s]}
	\psfrag{y} [B][B][0.7][0]{$u$ [Nm]}
	\includegraphics[clip = true, width = 0.23\textwidth]{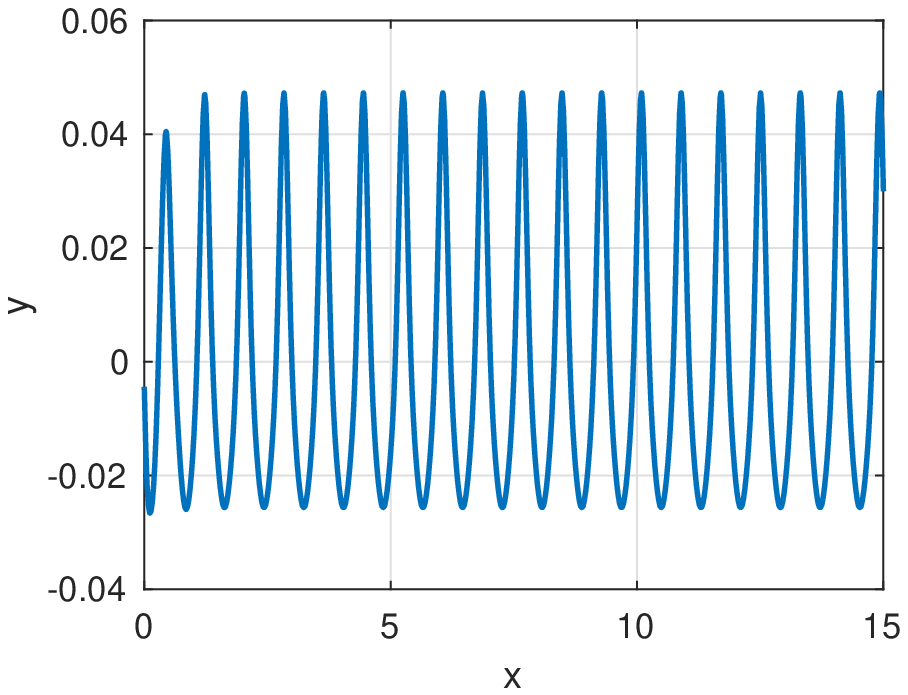}
	\hspace{0.001\textwidth}
	\psfrag{x} [B][B][0.7][0]{(g) time [s]}
	\psfrag{y} [B][B][0.7][0]{$d((x, \dot{x}), \Gamma_x)$}
	\includegraphics[clip = true, width = 0.23\textwidth]{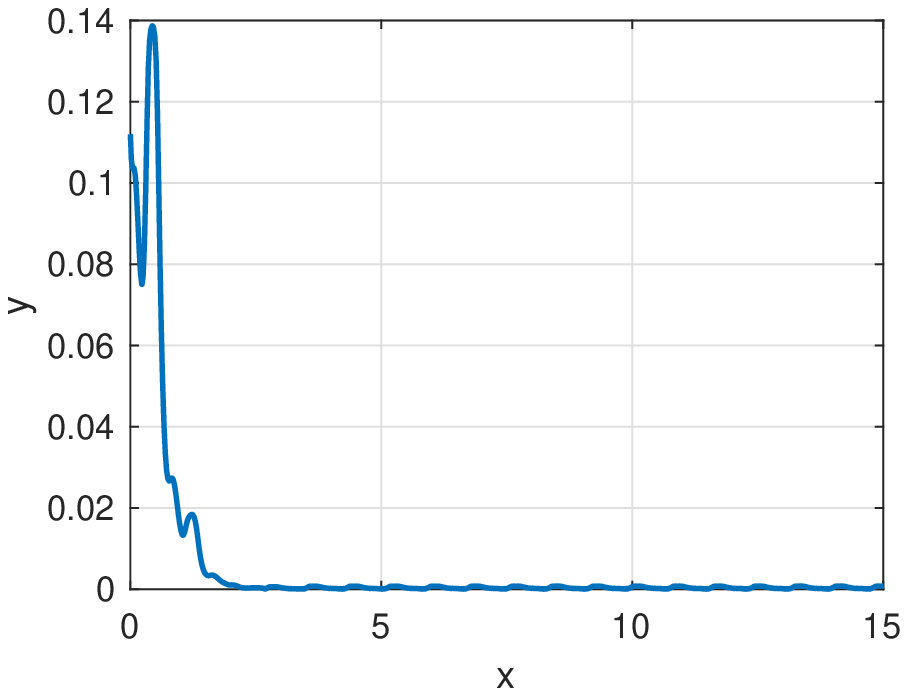}
	\hspace{0.001\textwidth}
	\psfrag{x} [B][B][0.7][0]{(h) time [s]}
	\psfrag{y} [B][B][0.7][0]{$q$ [rad]}
	\psfrag{a} [B][B][0.5][0]{ $q_1$}
	\psfrag{b} [B][B][0.5][0]{ $q_2$}
	\psfrag{c} [B][B][0.5][0]{ $q_3$}
	\psfrag{d} [B][B][0.5][0]{ $q_4$}
	\includegraphics[clip = true, width = 0.23\textwidth]{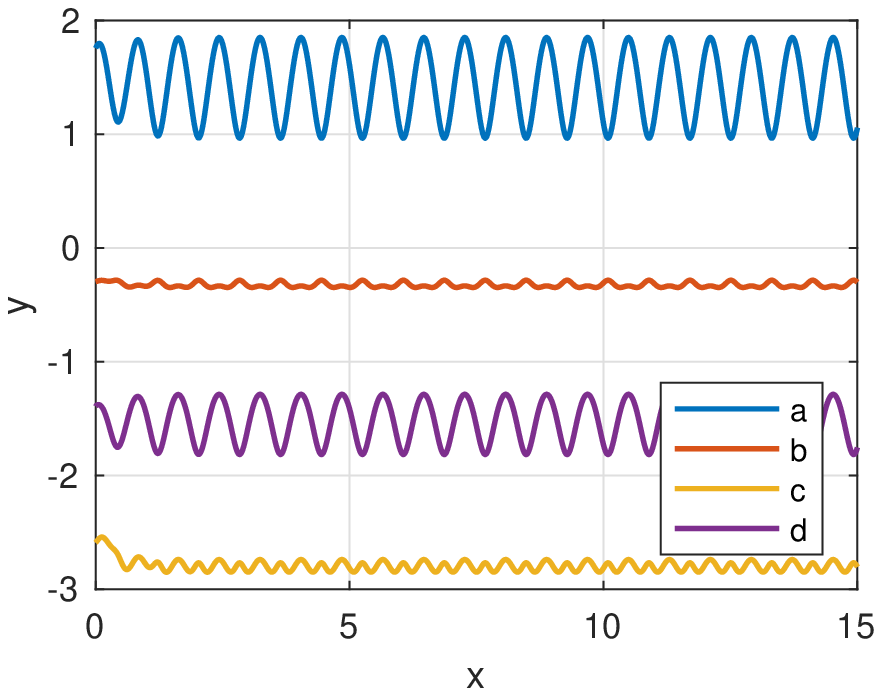}
	\end{subfigure}
		
	\vspace{3pt}
	
	\caption{Orbital stabilization results, with $k_{\text{b}} = 0.116 \text{Nm}/\text{rad}$, $k_{\text{t}} = 0 \text{Nm}/\text{rad}$, $\delta_{\text{b}} = 0.0183\text{m}$, $\delta_{\text{t}} = 0.016\text{m}$ (obtained from the 4-parameter optimization). (a): $y$ (blue) and reference $\bar{r}_y$ (red). (b): $\dot{y}$. (c): evolution of $(x, \dot{x})$ (blue) and $(\bar{x}, \dot{\bar{x}})$ (red). (d): periodic behavior of $x$. (e): tracking error $e$. (f): control input $u$. (e): normalized distance between $(x, \dot{x})$ and the orbit $\Gamma_x$. (f): behavior of the joint angles.}
	\label{fig:control_simulation2}
\end{figure}

 The results of the simulations are presented in Figure \ref{fig:control_simulation1}, where we employed the stiffness values $(0.1283, 0) \text{Nm/rad}$ (with $\delta_{\text{t}} = \delta_\text{b}  =0.02$m) obtained from the first optimization procedure, and in Figure \ref{fig:control_simulation2}, where we used the parameter set (0.116 \text{Nm}/\text{rad}, 0 \text{Nm}/\text{rad}, 0.0183 \text{m}, 0.016 \text{m}), obtained from the second optimization procedure.
For both simulation scenarios, we presented the output tracking performance and the internal dynamics behavior.
In particular, we highlighted the distance from the target orbit in the subplots (g), where we used as distance the following expression:
\begin{equation}
\begin{split}
d((x, \dot{x}), \Gamma_x) = \min_{(\xi_1, \xi_2) \in \Gamma_x}\left|\begin{pmatrix} \frac{x - \xi_1}{\Delta_1} \\ \frac{\dot{x} - \xi_2}{\Delta_2} \end{pmatrix} \right|, &\qquad \Gamma_x = \{(\xi_1, \xi_2) : \xi_1 = \bar{x}(t), \xi_2 = \dot{\bar{x}}(t), 0 \leq t \leq T_q \}\\
\Delta_1 = \max_{(\xi_1, \xi_2) \in \Gamma_x}\xi_1 - \min_{(\xi_1, \xi_2) \in \Gamma_x}\xi_1, &\qquad \Delta_2 = \max_{(\xi_1, \xi_2) \in \Gamma_x}\xi_2 - \min_{(\xi_1, \xi_2) \in \Gamma_x}\xi_2.
\end{split}
\end{equation}
We remark that we introduced the positive normalizing terms $\Delta_1$, $\Delta_2$ to highlight the transient of $(x, \dot{x})$ towards the target orbit.
From the presented results, the effectiveness of the controller is confirmed, and after a brief transient all signals reach the desired law of motion.
For completeness, the subplots (h) indicate the joint coordinates corresponding to the physical evolution of the mechanism.

\section{Conclusions}\label{sec:conclusions}
We presented an integrated design strategy for a 2-DOF underactuated mechanism, combining structural optimization and control.
Some parameters related to the elasticities and the mass distribution were employed to generate periodic trajectories in the system's zero dynamics.
Therefore, it was possible to optimize the system response in a synergistic fashion by combining the resulting oscillating behavior with a controller for orbital stabilization.
This preliminary work presents several directions for possible extension.
Among these, we find the application of the procedure to a general class of underactuated mechanisms, an increase of the decision variables to achieve a more complex behavior, and the inclusion of adaptive/constrained control techniques to ensure robustness in the presence of model uncertainties.

\bibliographystyle{ieeetr}
{\bibliography{synergistic_design}}\

\begin{thebibliography}{10}

\bibitem{wasfy2003computational}
T.~M. Wasfy and A.~K. Noor, ``Computational strategies for flexible multibody
  systems,'' {\em Appl. Mech. Rev.}, vol.~56, no.~6, pp.~553--613, 2003.

\bibitem{tromme2018system}
E.~Tromme, A.~Held, P.~Duysinx, and O.~Br{\"u}ls, ``System-based approaches for
  structural optimization of flexible mechanisms,'' {\em Archives of
  Computational Methods in Engineering}, vol.~25, no.~3, pp.~817--844, 2018.

\bibitem{ma2006multidomain}
Z.-D. Ma, N.~Kikuchi, C.~Pierre, and B.~Raju, ``Multidomain topology
  optimization for structural and material designs,'' {\em Journal of Applied
  Mechanics}, vol.~73, no.~4, pp.~565--573, 2006.

\bibitem{trease2009design}
B.~Trease and S.~Kota, ``Design of adaptive and controllable compliant systems
  with embedded actuators and sensors,'' {\em Journal of Mechanical Design},
  vol.~131, no.~11, 2001.

\bibitem{de2002underactuated}
A.~De~Luca, ``Underactuated manipulators: control properties and techniques,''
  {\em Machine Intelligence and Robotic Control}, vol.~4, no.~3, pp.~113--126,
  2002.

\bibitem{liu2013survey}
Y.~Liu and H.~Yu, ``A survey of underactuated mechanical systems,'' {\em IET
  Control Theory \& Applications}, vol.~7, no.~7, pp.~921--935, 2013.

\bibitem{olfati2001nonlinear}
R.~Olfati-Saber, {\em Nonlinear control of underactuated mechanical systems
  with application to robotics and aerospace vehicles}.
\newblock PhD thesis, Massachusetts Institute of Technology, 2001.

\bibitem{spong1994partial}
M.~W. Spong, ``Partial feedback linearization of underactuated mechanical
  systems,'' in {\em Proceedings of IEEE/RSJ International Conference on
  Intelligent Robots and Systems (IROS'94)}, vol.~1, pp.~314--321, IEEE, 1994.

\bibitem{ortega2002stabilization}
R.~Ortega, M.~W. Spong, F.~G{\'o}mez-Estern, and G.~Blankenstein,
  ``Stabilization of a class of underactuated mechanical systems via
  interconnection and damping assignment,'' {\em IEEE Transactions on Automatic
  Control}, vol.~47, no.~8, pp.~1218--1233, 2002.

\bibitem{shiriaev2005constructive}
A.~Shiriaev, J.~W. Perram, and C.~Canudas-de Wit, ``Constructive tool for
  orbital stabilization of underactuated nonlinear systems: Virtual constraints
  approach,'' {\em IEEE Transactions on Automatic Control}, vol.~50, no.~8,
  pp.~1164--1176, 2005.

\bibitem{mohammadi2018dynamic}
A.~Mohammadi, M.~Maggiore, and L.~Consolini, ``Dynamic virtual holonomic
  constraints for stabilization of closed orbits in underactuated mechanical
  systems,'' {\em Automatica}, vol.~94, pp.~112--124, 2018.

\bibitem{seifried2014dynamics}
R.~Seifried, {\em Dynamics of underactuated multibody systems}.
\newblock Springer, 2014.

\bibitem{bastos2019synergistic}
G.~Bastos, ``A synergistic optimal design for trajectory tracking of
  underactuated manipulators,'' {\em Journal of Dynamic Systems, Measurement,
  and Control}, vol.~141, no.~2, 2019.

\bibitem{bosso2019constrained}
A.~Bosso, A.~Serrani, C.~Conficoni, and A.~Tilli, ``Constrained-inversion
  {MRAC}: An approach combining hard constraints and adaptation in uncertain
  nonlinear systems,'' in {\em 2019 IEEE 58th Conference on Decision and
  Control (CDC)}, pp.~2039--2045, IEEE, 2019.

\bibitem{sreenath2011compliant}
K.~Sreenath, H.-W. Park, I.~Poulakakis, and J.~W. Grizzle, ``A compliant hybrid
  zero dynamics controller for stable, efficient and fast bipedal walking on
  mabel,'' {\em The International Journal of Robotics Research}, vol.~30,
  no.~9, pp.~1170--1193, 2011.

\bibitem{khalil_3ed}
H.~K. Khalil, {\em Nonlinear Systems, Third Edition}.
\newblock Prentice-Hall, 2002.

\bibitem{shiriaev2006periodic}
A.~Shiriaev, A.~Robertsson, J.~Perram, and A.~Sandberg, ``Periodic motion
  planning for virtually constrained {E}uler--{L}agrange systems,'' {\em
  Systems \& Control Letters}, vol.~55, no.~11, pp.~900--907, 2006.

\end{thebibliography}

\end{document}